\documentclass[showpacs,aps,twocolumn]{revtex4}
\usepackage{bm}
\usepackage{amsmath}
\usepackage{graphicx}
\usepackage{subfigure}
\usepackage[usenames,dvipsnames]{color}
\definecolor{darkblue}{RGB}{0,0,196}
\usepackage[colorlinks=true,linkcolor=darkblue,citecolor=darkblue,urlcolor=darkblue]{hyperref}
\usepackage{setspace}
\usepackage{hyperref}
\usepackage{footmisc}
\usepackage[makeroom]{cancel}
\usepackage{comment}
\usepackage{lineno}
\def\be{\begin{equation}}
\def\ee{\end{equation}}
\def\ba{\begin{eqnarray}}
\def\ea{\end{eqnarray}}
\usepackage{graphicx}
\usepackage{amsmath,bbm}
\usepackage{amssymb,bm}
\usepackage{yfonts}

\begin{document}
\title{Event Shape Engineering and Multiplicity dependent Study of Identified Particle Production in Proton+Proton Collisions at $\sqrt{s}$ = 13 TeV using PYTHIA8}
\author{Arvind Khuntia}
\author{Sushanta Tripathy}
\author{Ashish Bisht}
\author{Raghunath Sahoo\footnote{Corresponding Author Email: Raghunath.Sahoo@cern.ch}}
\affiliation{Department of Physics, Indian Institute of Technology Indore, Simrol, Indore 453552, India}

\begin{abstract}
\noindent
Small system collectivity observed at the LHC energies along with enhancement of strangeness makes high-multiplicity proton+proton (pp) collisions very interesting in order to look for QGP-like features, usually found in heavy-ion collisions. It may be interesting to perform a double differential study of different observables in pp collisions in terms of charged particle multiplicity and event shape in order to understand the new dimensions in high-multiplicity pp physics. We study the correlation between the number of multi-partonic interactions (nMPI), event shape (transverse spherocity) and charged particle multiplicity classes. For the first time, we report the simulation results on the spherocity and charged particle multiplicity dependent study of ($\pi^{+}+\pi^{-}$), (K$^{+}$+K$^{-}$), (p+$\mathrm{\bar{p}}$), K$^{*0}$, $\phi$ and ($\Lambda+\bar{\Lambda}$) production in pp collisions at $\sqrt{s}$ = 13 TeV using PYTHIA8. We explore the event shape and charged particle multiplicity dependence of the transverse momentum ($p_{\rm{T}}$) spectra, integrated yield, mean transverse momentum ($\langle p_{\rm{T}} \rangle$) and particle ratios of the identified particles. This study provides a baseline for exploring the the event topology and final state multiplicity dependence of identified particle production in the LHC pp collisions.

\end{abstract}
\date{\today}
\maketitle 

\section{Introduction}
\label{intro}
Recent multiplicity dependent measurements of identified particle production from the experiments at the Large Hadron Collider (LHC)~\cite{ALICE:2017jyt} have revealed QGP-like behaviour in high-multiplicity proton+proton (pp) collisions. These raise concerns whether pp collisions can be used as a proper benchmark for comparison with heavy-ion results to understand the formation of a medium with high temperature/energy density. Recently, there have been attempts to look into possible QGP-like signature in central pp collisions \cite{Mangano:2017plv}. Such behaviours have important consequences in understanding the data from heavy-ion collisions at the LHC energies as one should consider the contribution of QGP-like effects in small systems. 
The origin of these effects needs proper investigation as it has been shown that hydro calculations~\cite{Bzdak:2014dia,Bozek:2013ska}, where a hot and dense QCD medium is explicitly assumed, can describe many features of experimental data. However, signature like long-range two-particle azimuthal correlation is qualitatively explained by the initial state effects \cite{Dumitru:2010iy}, whereas an incoherent elastic scattering of partons as discussed in AMPT also explains this feature \cite{Ma:2014pva}. PYTHIA8 \cite{Sjostrand:2007gs} seems to reproduce collectivity-like features in pp collisions, which are attributed to the multi-parton interactions (MPI) \cite{Sjostrand:2013cya} and color reconnection (CR) \cite{Ortiz:2013yxa} mechanisms. Because of the composite nature of protons, it is possible to have events with multiple parton-parton interactions (MPI) in a single pp collision. This process is one of the key ingredients in PYTHIA8~\cite{Sjostrand:2006za} and it is well supported by experimental data in $\mathrm{p\bar{p}}$ collisions~\cite{Abazov:2009gc,Chekanov:2007ab}. The linear increase of J/$\psi$ production with multiplicity \cite{Abelev:2012rz} at the forward rapidities in pp collisions at the LHC is very well explained through the MPI and CR mechanisms in PYTHIA8 \cite{Thakur:2017kpv}. The parameters for MPI are usually tuned by looking at the observables like charged particle multiplicity, transverse momentum ($p_{\rm{T}}$) and their correlations. But, this approach excludes many details of pp interactions, which leads to failure of describing different observables like the strange particle production~\cite{Abelev:2012hy,Aamodt:2011zza} and jet production rate~\cite{Abelev:2012sk} as a function of the event multiplicity. Thus, one should look into new observables which will help to understand the component of hadronic interactions - hard (pQCD) or the soft, which fails to be described by theory, eventually causing overall disagreements. From PYTHIA 8.180 onwards, the CR mechanism is implemented which explains flow-like effects in pp collisions \cite{Ortiz:2013yxa}. These effects increase with the number of MPI and the average MPI increases with event multiplicity but it saturates for high-multiplicities~\cite{Ortiz:2013yxa,Cuautle:2014yda,Cuautle:2015kra}. The major differences in underlying models in PYTHIA8 arise from the way mesons and baryons are formed in String Fragmentation model (SFM). In SFM, as the quark and anti-quark move apart, potential energy stored in the string increases and the string may break by producing a new quark-antiquark pair. Eventually, it produces mesons. For the production of baryons, diquark in a color anti-triplet state is treated like an ordinary antiquark, in such a manner that a string can break either by quark-antiquark or anti-diquark-diquark pair production. The tunnelling mechanism in SFM implies a suppression of heavy-quark production, $u:d:s:c \approx 1:1:0.3:10^{-11}$, which creates the differences in the production of strange and non-strange particles. In Refs.~\cite{Bierlich:2015rha,Acconcia:2017bjv}, it is shown that color reconnection mechanism also modifies the final string fragmentation into hadrons and in particular, an enhancement of baryons over meson is observed with increasing multiplicity. Event shape observables allow the possibility to separate the high  and low number of MPI events to isolate the behaviour of particles inside jets (hard processes) and that pertaining to the bulk (soft processes).  It has also been reported in Refs.~\cite{Cuautle:2014yda,Cuautle:2015kra} that, the average number of MPI ($<$MPI$>$) increases with the spherocity i.e. low $<$MPI$>$ for jetty events while high $<$MPI$>$ for isotropic events.
\begin{figure}[ht]
\includegraphics[width=8.5cm, height=7.0cm]{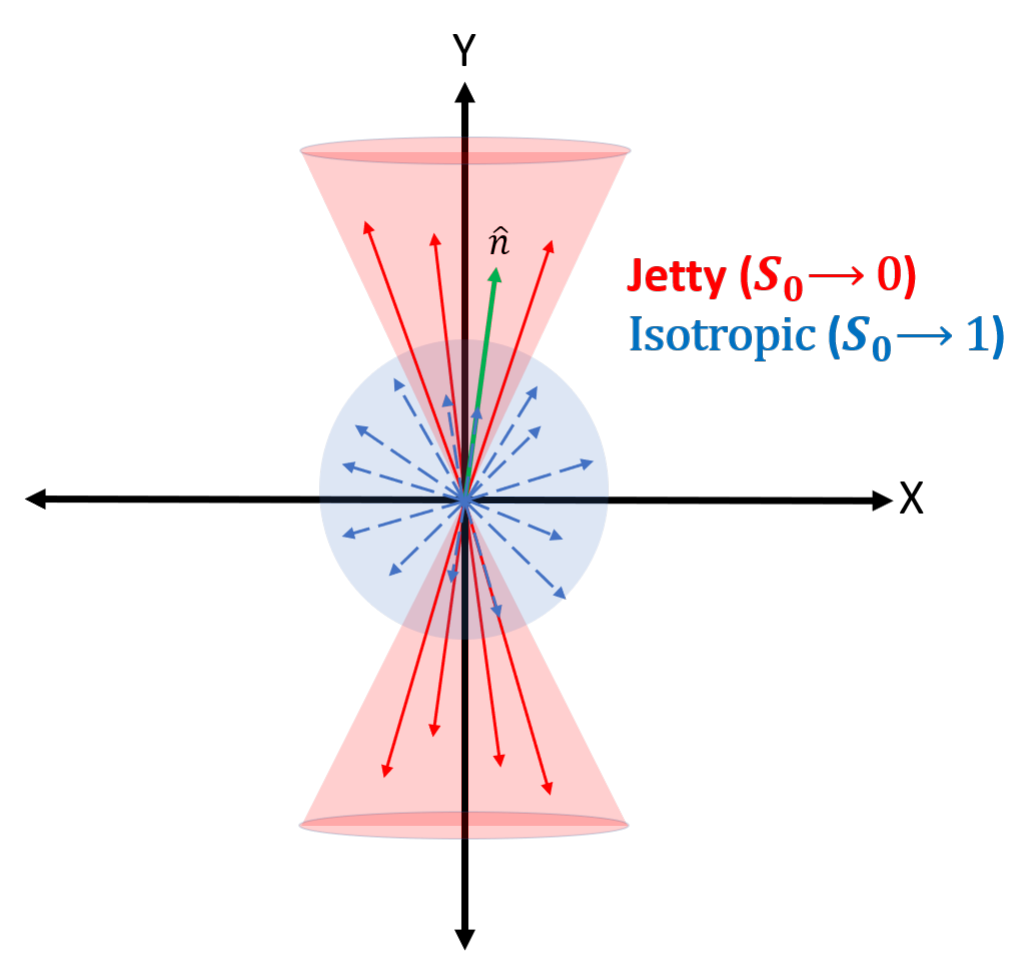}
\caption[]{(Color Online) Schematic picture showing jetty and isotropic events in the transverse plane.}
\label{sp_cart}
\end{figure}

Event shape studies at the LHC~\cite{Abelev:2012sk} allows to extract more information from the data by separating the jetty (high $p_{\rm{T}}$-jets) and isotropic events (low-$Q^2$ partonic scatterings). Most of the event generators reported in Ref.~\cite{Abelev:2012sk} overestimate the contribution of events with back-to-back jet structure while underestimating the contribution of isotropic events at high-multiplicity ($N_{\rm ch} > 30$). This suggests that the average measurements do not contain enough information and proper care is needed while extracting physics information from models/event generators. To understand the influence of multi-partonic interactions on the final state in pp collisions, the study of event shape (event topology) has to be performed in data and models as well. It has been reported that in the transverse spherocity event classes, the distribution of the number of multi-partonic interactions (nMPI) are narrower compared to those without any selection of the event shape~\cite{Cuautle:2015kra}. This result can help to understand more on the jet production, identified particle ratios (baryon to meson and strange to non-strange) and the steep rise of mean transverse momenta of charged particles in small systems. A comprehensive differential study using event shapes would reveal interesting features, which could be exploited to get physical information as well as to improve theoretical models in the MC event generators. In Ref.~\cite{Bencedi:2018ctm}, the preliminary results from ALICE in pp collisions at $\sqrt{s}$ = 13 TeV as a function of event shape and charged particle multiplicity are compared with event generators such as PYTHIA8 and EPOS-LHC. The event generators seem to describe the data qualitatively. 

In this paper, we perform a comprehensive double differential (event shape and multiplicity) study of identified particle production in terms of $p_{\rm{T}}$-spectra, integrated yield, mean $p_{\rm{T}}$ and the $p_{\rm{T}}$-differential and $p_{\rm{T}}$-integrated particle ratios. We also explore the dependence of event shape and charged particle multiplicity on the number of multi-partonic interactions in pp collisions. Although PYTHIA as an event generator has different tunes (for incorporating various physics processes), for the present study, we focus on a specific tune which includes MPI and CR (MONASH tune). As a limitation, like any other theoretical models, we do not compare the effect of
different physics processes, which becomes out of the scope of the present study.

This paper is organized as follows. We start with a complete description of the event generation in PYTHIA8 and the analysis methodology in Section~\ref{ev_ana}. The correlation of MPI, charged particle multiplicity and transverse spherocity are described  in Section~\ref{sp_mpi}.  In Section~\ref{results}, we discuss the results such as the $p_{\rm{T}}$-spectra, integrated yield, mean $p_{\rm{T}}$ and the particle ratios for identified particles. Finally in Section~\ref{summary}, we summarize the
work with important findings. 

\section{Event Generation and Analysis Methodology}
\label{ev_ana}
PYTHIA is an event generator used to simulate ultra-relativistic collision events at high energies taking $e^{\pm}, $~p$,\rm{~and~} \bar{p} $. It is a pQCD based model with physics processes like hard and soft interactions, parton distributions, initial- and final-state parton showers, multipartonic interactions, fragmentation, color reconnection and decay~\cite{Sjostrand:2006za}.

In our present study, we have used PYTHIA 8.235, an advanced version of PYTHIA6 which includes the multi-partonic interaction (MPI) scenario as one of the key improvements. In PYTHIA, MPI scenario is crucial to explain the underlying events, multiplicity distributions and flow-like patterns in terms of color reconnection. A detailed explanation of the physics processes in PYTHIA 8.235 can be found in Ref.~\cite{pythia8html}. We have implemented the inelastic, non-diffractive component of the total cross-section for all soft QCD processes (SoftQCD : all = on). This analysis is carried out with 250 million events at $ \sqrt{s}=13~\mathrm{TeV} $ with Monash 2013 Tune (Tune:14)~\cite{Skands:2014pea} and MPI based scheme of color reconnection (ColorReconnection:mode(0)). For the generated events, we let all the resonances to decay except the ones used in our study (HadronLevel:Decay = on). The event selection criteria throughout the analysis is such that only those events are chosen which have at least 5 tracks (charged particles) at the mid-rapidity ($|\eta| < 0.8$). Charged particle multiplicities have been chosen in the acceptance of V0 detector in ALICE at the LHC with pseudo-rapidity coverage of V0A ($2.8<\eta<5.1$) and V0C ($-3.7<\eta<-1.7$)~\cite{Abelev:2014ffa}. These events are categorized in ten V0 multiplicity (V0M) bins and we define V0M-I as the top 10 percent of events and V0M-X as the lowest 10 percent of events. The charged particle multiplicities in each event in different V0M classes (all the events are divided into multiplicity classes based on the total charge deposited in both V0 detectors, called the “V0M amplitude") are listed in Table~\ref{tab:V0M}. The minimum bias events are those events where no selection on charged particle multiplicity is applied. To disentangle the jetty and isotropic events from the average-shaped events, we have applied spherocity (defined in next section) cuts on generated events. In this analysis, the spherocity distributions are selected in the pseudo-rapidity range of $|\eta|<0.8$ with a minimum constraint of 5 charged particles with $p_{\rm{T}}~ $ $>$ 0.15 GeV/$c$. The jetty events are those having $0\leq S_{0}<0.29$ with lowest 20 percent and the isotropic events are those having $0.64<S_{0}\leq1$ with highest 20 percent of the total minimum bias events~\cite{Bencedi:2018ctm}. In this work, we use
a shorthand notation $N_{\rm ch}$ for the charged particle multiplicity in the V0 detector acceptance.

\begin{table}[h]
\caption{V0M multiplicity classes and the corresponding charged particle multiplicities.}
\centering 
\scalebox{0.8}
{
\begin{tabular}{|c|c|c|c|c|c|c|c|c|c|c|} 
\hline 
V0M class & I & II & III & IV & V & VI & VII &VIII & IX & X \\
\hline 
$N_{\rm ch}$ &50-140 & 42-49 & 36-41 & 31-35 & 27-30 & 23-26 & 19-22 &  15-18 & 10-14 & 0-9\\
\hline
\end{tabular}
}
\label{tab:V0M}
\end{table}

\noindent
\section{Transverse spherocity, Multi-partonic interactions and Charged particle multiplicity}
\label{sp_mpi}
Transverse spherocity for an event is defined for a unit vector $\hat{n} (n_{T},0)$ which minimizes $S_0$~\cite{Cuautle:2014yda, Cuautle:2015kra, Ortiz:2017jho}:
\begin{eqnarray}
S_{0} = \frac{\pi^{2}}{4} \bigg(\frac{\Sigma_{i}~|\vec p_{T_{i}}\times\hat{n}|}{\Sigma_{i}~p_{T_{i}}}\bigg)^{2}.
\end{eqnarray}

By restricting it to transverse plane, spherocity becomes infrared and collinear safe~\cite{Salam:2009jx}. By construction, the extreme limits of spherocity are related to specific configurations of events in the transverse plane. The limit of spherocity is in between 0 and 1. Spherocity becoming 0 would mean that the events are pencil-like (back-to-back structure) while 1 would mean the events are isotropic. The pencil-like events are hard events while the isotropic events are the result of soft processes. Figure~\ref{sp_cart} depicts the jetty and isotropic events in the transverse plane. 

\begin{figure}[ht!]
\includegraphics[width=8.cm, height=7.cm]{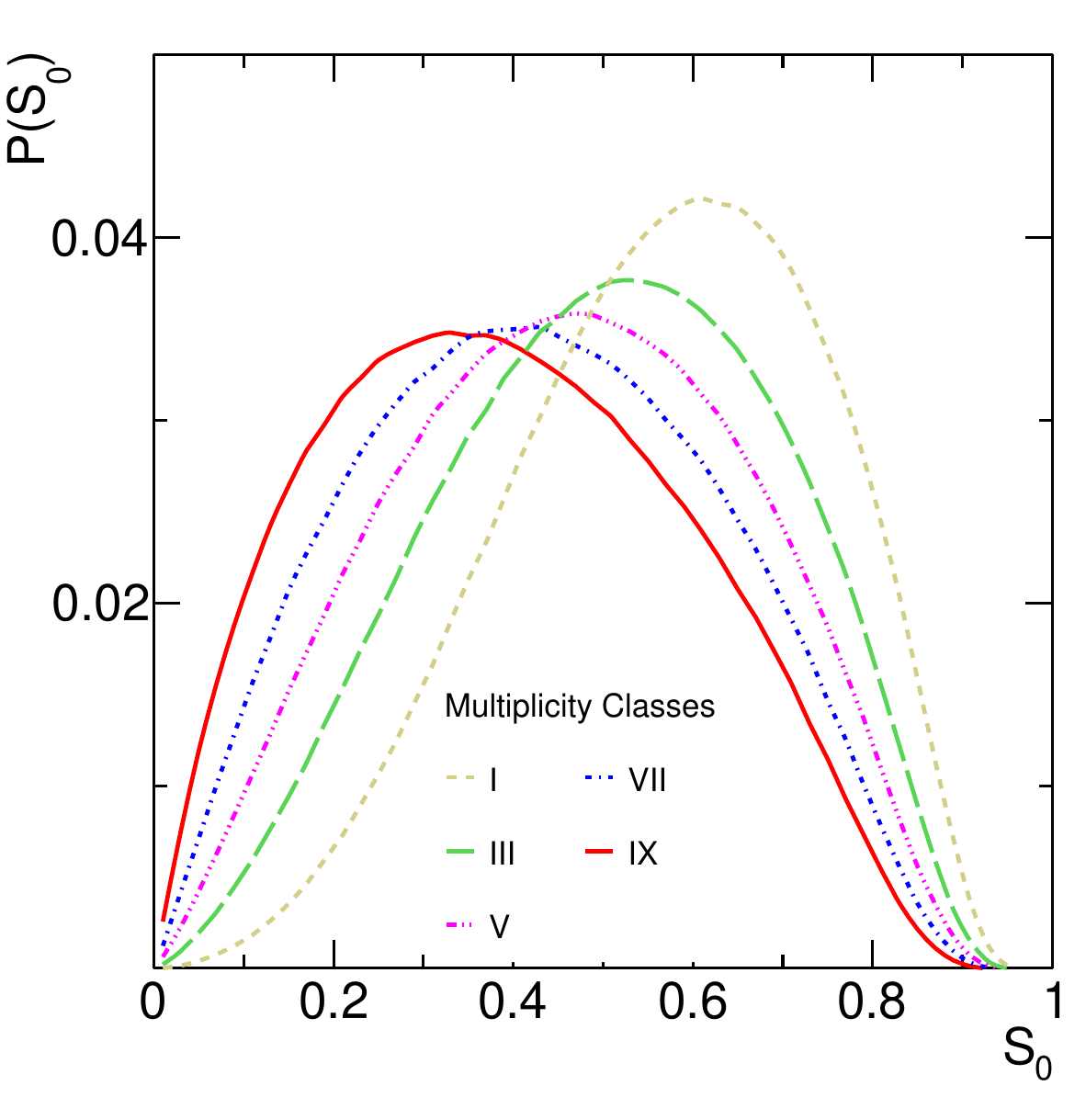}
\includegraphics[width=8.cm, height=7.cm]{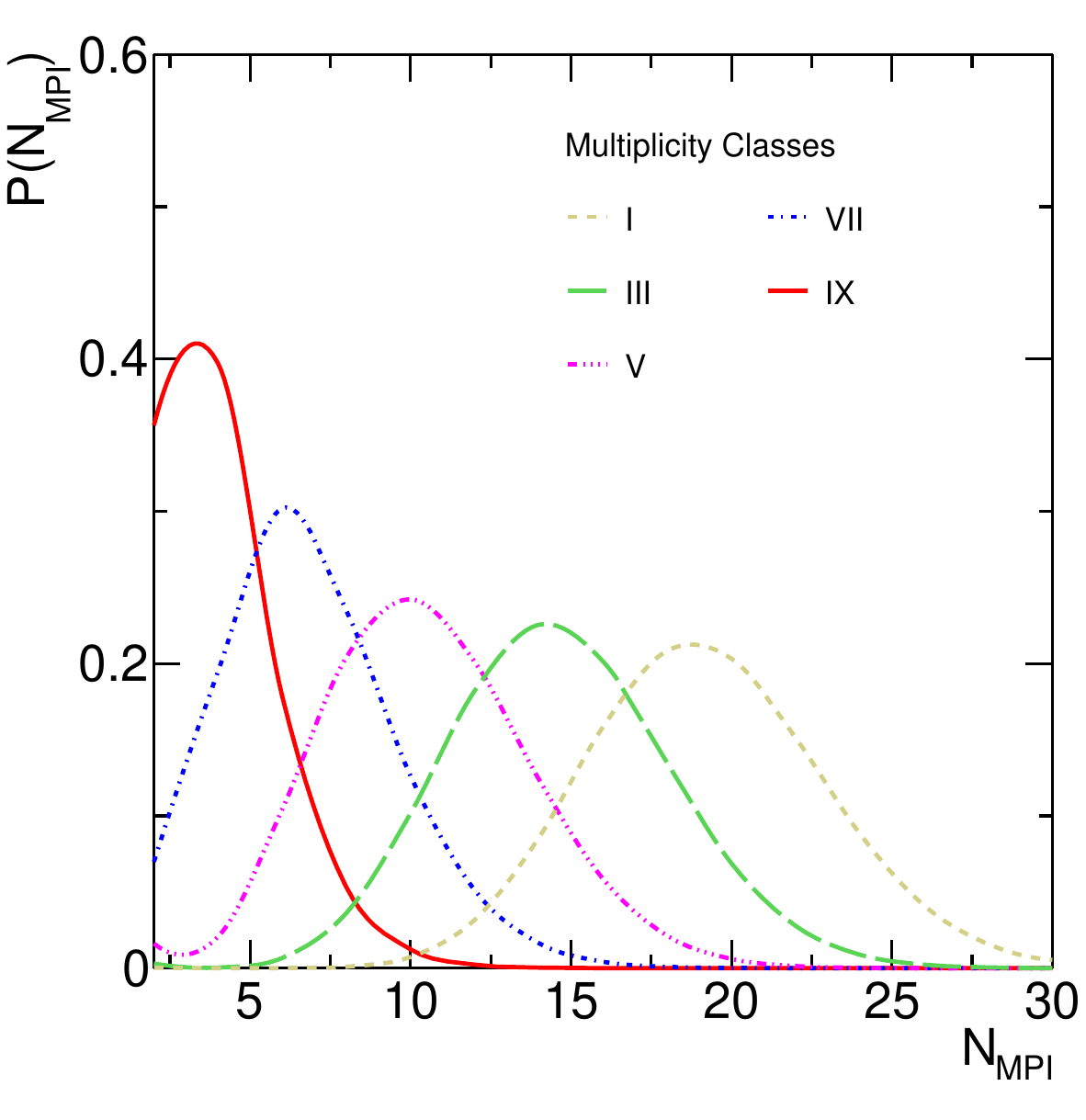}
\caption[]{(Color Online) Spherocity (upper panel) and nMPI (lower Panel) distributions for different charged particle multiplicity classes in pp collisions at $ \sqrt{s} =\mathrm{13~TeV}$. Different line styles and colors are for different multiplicity classes. }
\label{sp_dis}
\end{figure}

Upper(Lower) panel of Fig.~\ref{sp_dis} shows the correlation between spherocity(nMPI) and charged particle multiplicity. A clear charged particle multiplicity dependence of spherocity and nMPI distributions is observed. As we move from low to high-multiplicity events, the peak of the spherocity distribution shifts towards right. This indicates that the high-multiplicity pp collisions are dominated by isotropic events while the low-multiplicity events are dominated by the jetty ones. The nMPI distributions suggest that large number of multi-partonic interactions occur for high-multiplicity pp collisions. From Fig.~\ref{sp_dis}, one observes that the width of the nMPI distribution increases from lower to higher multiplicity classes making it squat, whereas the height goes down. For all multiplicity classes the distribution seems to be positively skewed. For a given class of multiplicity,  nMPI follows a distribution and towards high-multiplicity it shows a saturation behaviour \cite{Cuautle:2015kra}. Hence, 
event multiplicity and nMPI can't be used uniquely to classify events in pp collisions. We observe larger number of average nMPIs for isotropic than the jetty events. It is evident from the event shape analysis that spherocity along with the charged particle multiplicity (which is correlated with nMPI) should be preferred for a better selectivity of events. We use the spherocity distribution as shown in Fig.~\ref{sp_dis} to make a clear distinction between isotropic and jetty events. Further combining spherocity with event multiplicity, we study various observables as discussed in the following section to understand the dynamics of particle production in pp collisions at $\sqrt{s}$ = 13 TeV.
\begin{figure}[ht]
\includegraphics[width=8.0cm, height=9.0cm]{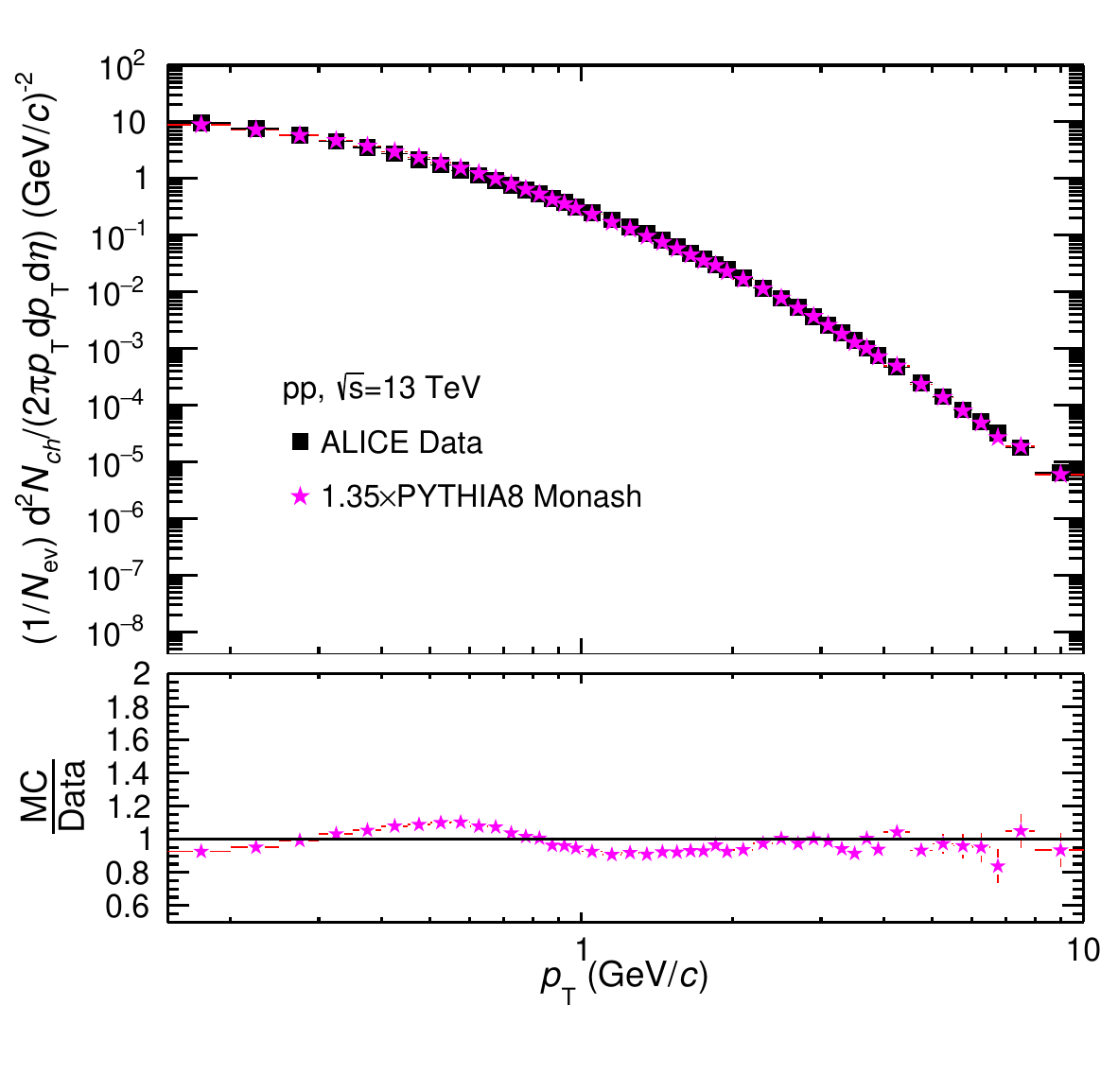}
\caption[]{(Color Online) Upper Panel: Comparison of charged particle $p_{\rm T}$ spectra in pp collisions at $\sqrt s$ = 13 TeV between ALICE data \cite{Adam:2015pza} and PYTHIA8 simulation. Lower Panel: the ratio between scaled simulated data and experimental data.}
\label{datVsPH}
\end{figure}

\section{Results and Discussion}
\label{results}
The first step in this analysis is to choose a proper PYTHIA8 tune to make the spectral shape compatible with experimental 
data. In order to do that, we have compared the charged particle $p_{\rm T}$ spectra for pp collisions at $\sqrt{s}$ = 13 TeV from ALICE data \cite{Adam:2015pza} and PYTHIA8 simulation in the same kinematic acceptance. This is shown in Fig. \ref{datVsPH}. The lower panel shows the ratio of the estimations from PYTHIA8 to experimental data. In order to see the agreement of spectral shapes, we have used an arbitrary scaling factor (1.35) to scale the simulated data in Fig. \ref{datVsPH}. We observe that the scaled simulated data agree with the spectral shape of the experimental data within (10-20)\% at low-$p_{\rm T}$ and consistent with unity for intermediate and high-$p_{\rm T}$.

As discussed in the previous section, we use spherocity as a tool to distinguish the isotropic and jetty events for each multiplicity class. We study the $p_{\rm{T}}$-spectra, integrated yield, mean transverse momentum for $(\pi^{+}+\pi^{-}),(\rm{K}^{+}+\rm{K}^{-}),(\rm{K}^{*0}+\rm{\overline{K}^{*0}})/2,(p+\bar{p}),\phi,(\Lambda^{0}+\bar{\Lambda}^{0})$ as a function of spherocity and charged particle multiplicity. We also study the $p_{\rm{T}}$-differential and $p_{\rm{T}}$-integrated particle ratios to $(\pi^{+}+\pi^{-})$ and $(\rm{K}^{+}+\rm{K}^{-})$, and $(p+\bar{p})$ to $\phi$ ratio as a function of spherocity for high-multiplicity pp collisions. From here onwards, $(\pi^{+}+\pi^{-}), (\rm{K}^{+}+\rm{K}^{-}),~(p+\bar{p}), (\rm{K}^{*0}+\rm{\overline{K}^{*0}})/2~ \rm{and} ~(\Lambda^{0}+\bar{\Lambda}^{0})$ are denoted as pion ($\pi$), kaon (K), proton (p), $\rm{K}^{*0}$ and $\Lambda$, respectively.

\subsection{$p_{\rm{T}}$-spectra}

\begin{figure}[ht]
\includegraphics[width=8.5cm, height=11.0cm]{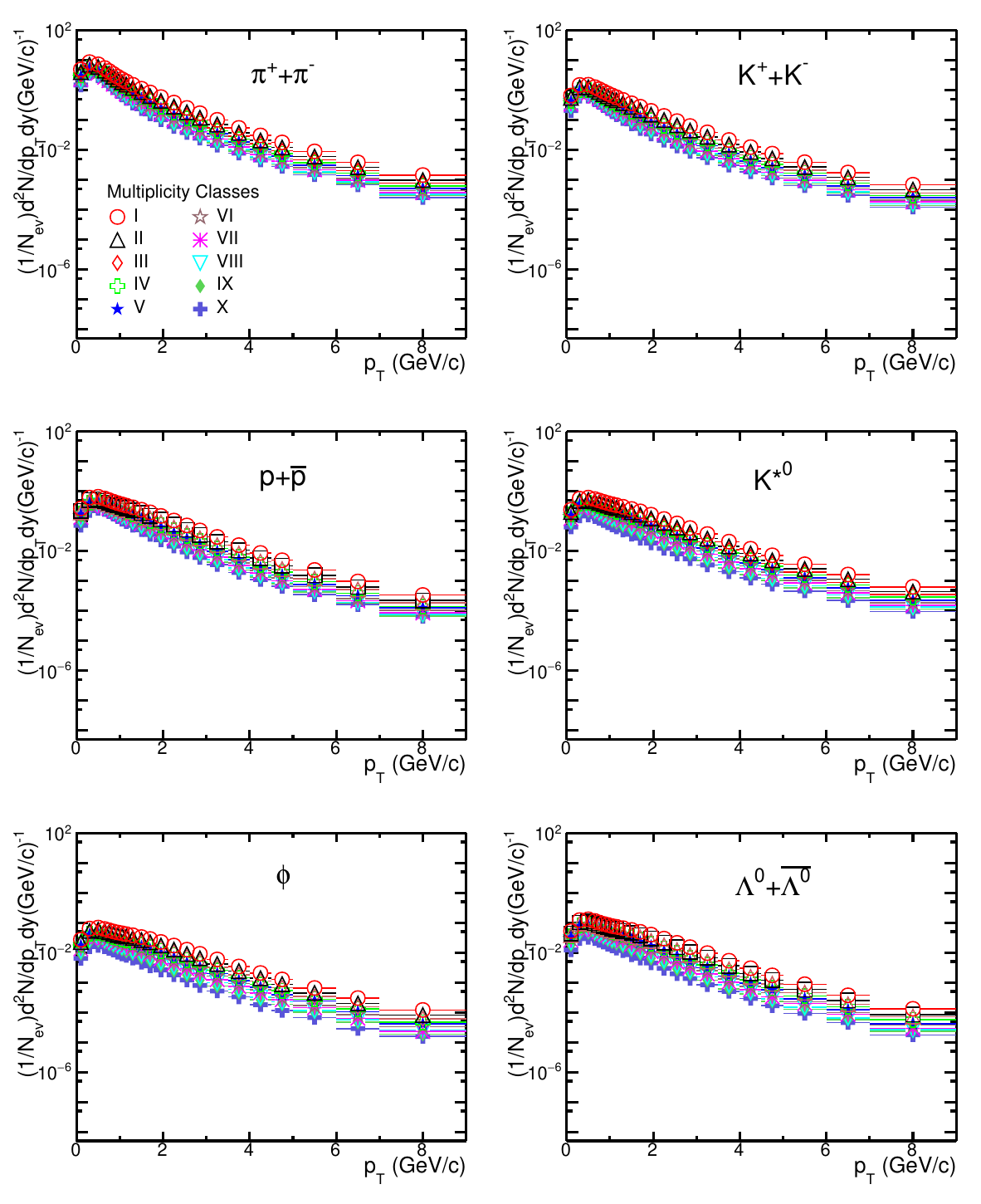}
\caption[]{(Color online) $p_{\rm{T}}$-spectra for light-flavor hadrons like pion, kaon, ${K}^{*0}$, proton, $\phi$, and $\Lambda$ at mid-rapidity ($ |\eta|< 0.5$) as a function of charged particle multiplicity for pp collisions at $\sqrt{s}=13\rm{~TeV}$ using PYTHIA8. Different markers and colors correspond to different multiplicity classes.}
\label{ptspectra_mult}
\end{figure}

\begin{figure*}[ht]
\includegraphics[width=8.5cm, height=9.5cm]{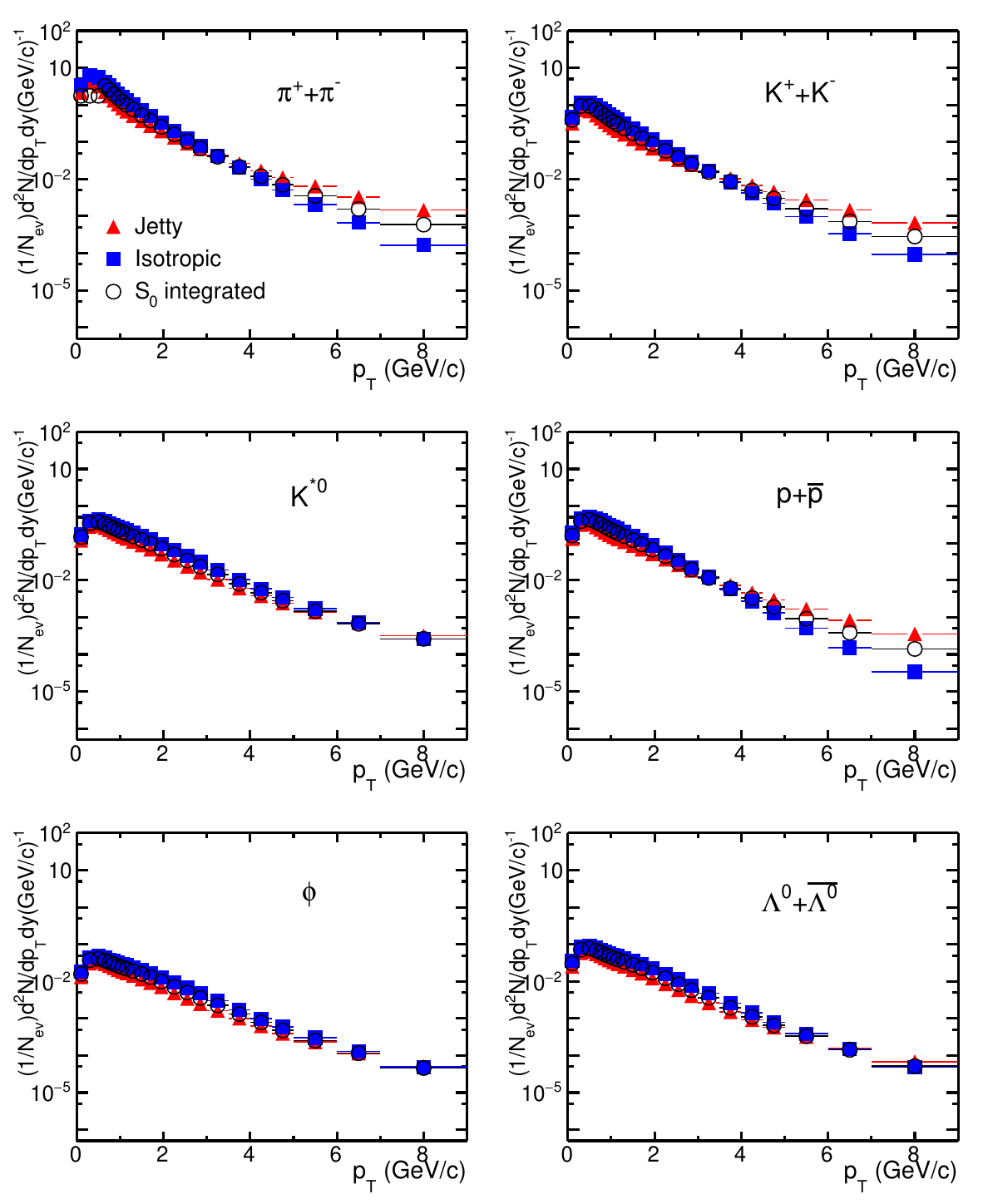}
\includegraphics[width=8.5cm, height=9.5cm]{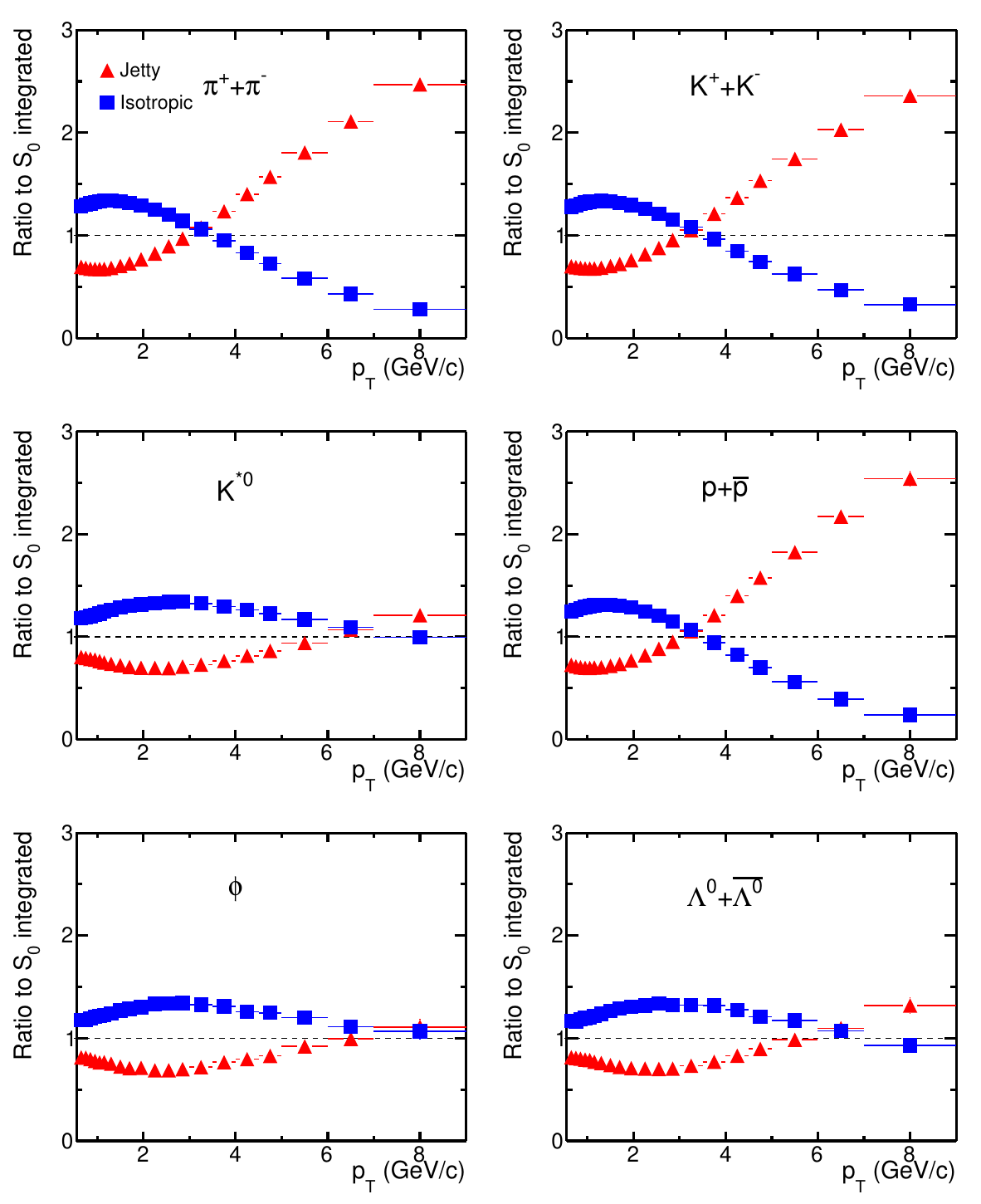}
\caption[]{(Color online) Left Panel: $p_{\rm{T}}$-spectra for pion, kaon, ${K}^{*0}$, proton, $\phi$, and $\Lambda$ for minimum bias pp collisions as a function of spherocity using PYTHIA8. The blue squares are for isotropic events, red triangles are for jetty events and open circles are for $S_{0}$ integrated events. Right Panel: Ratio of $p_{\rm{T}}$-spectra for isotropic and jetty events to the spherocity integrated events.}
\label{ptspectra_sp_min_bias}
\end{figure*}

\begin{figure*}[ht]
\includegraphics[width=8.5cm, height=9.5cm]{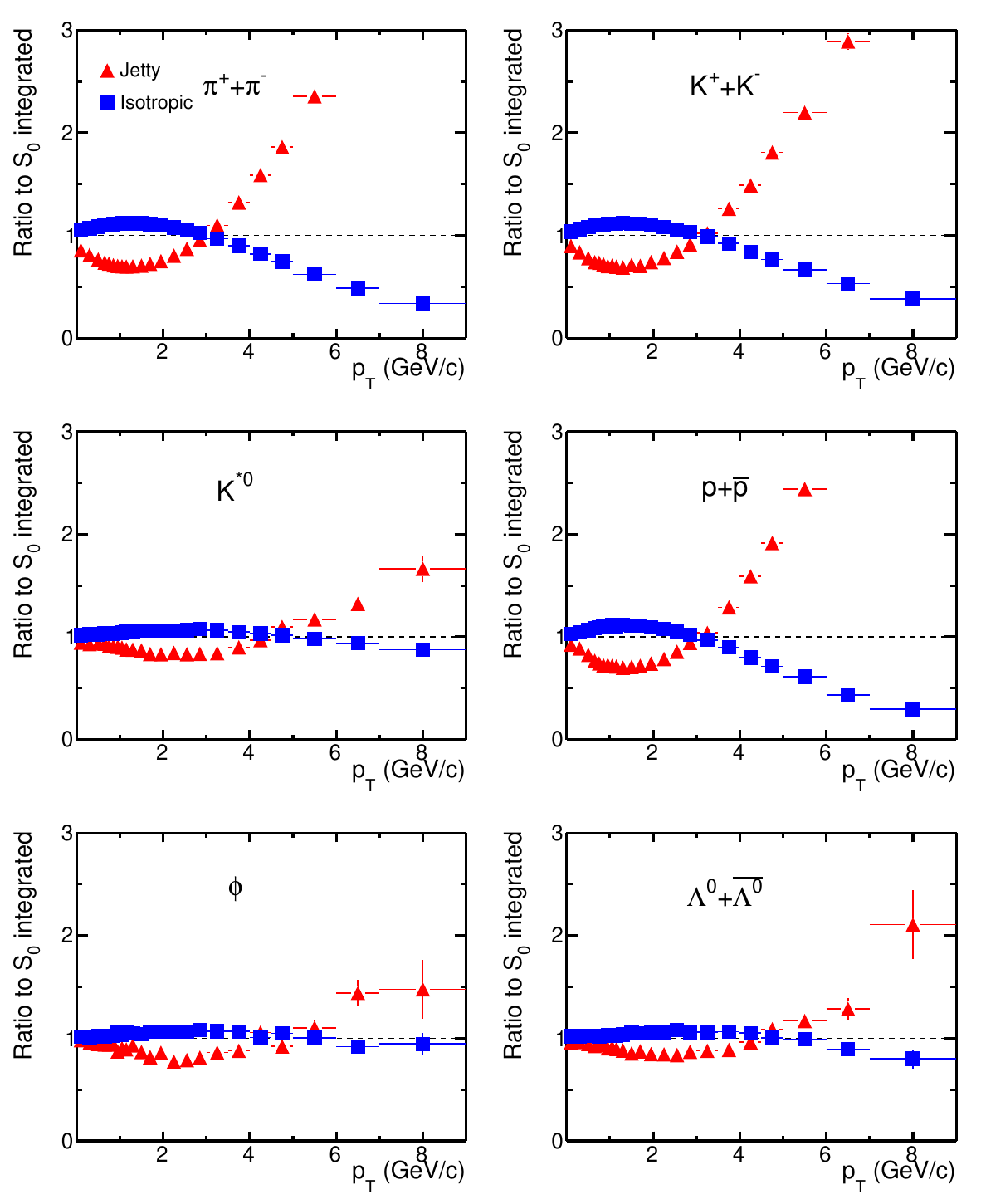}
\includegraphics[width=8.5cm, height=9.5cm]{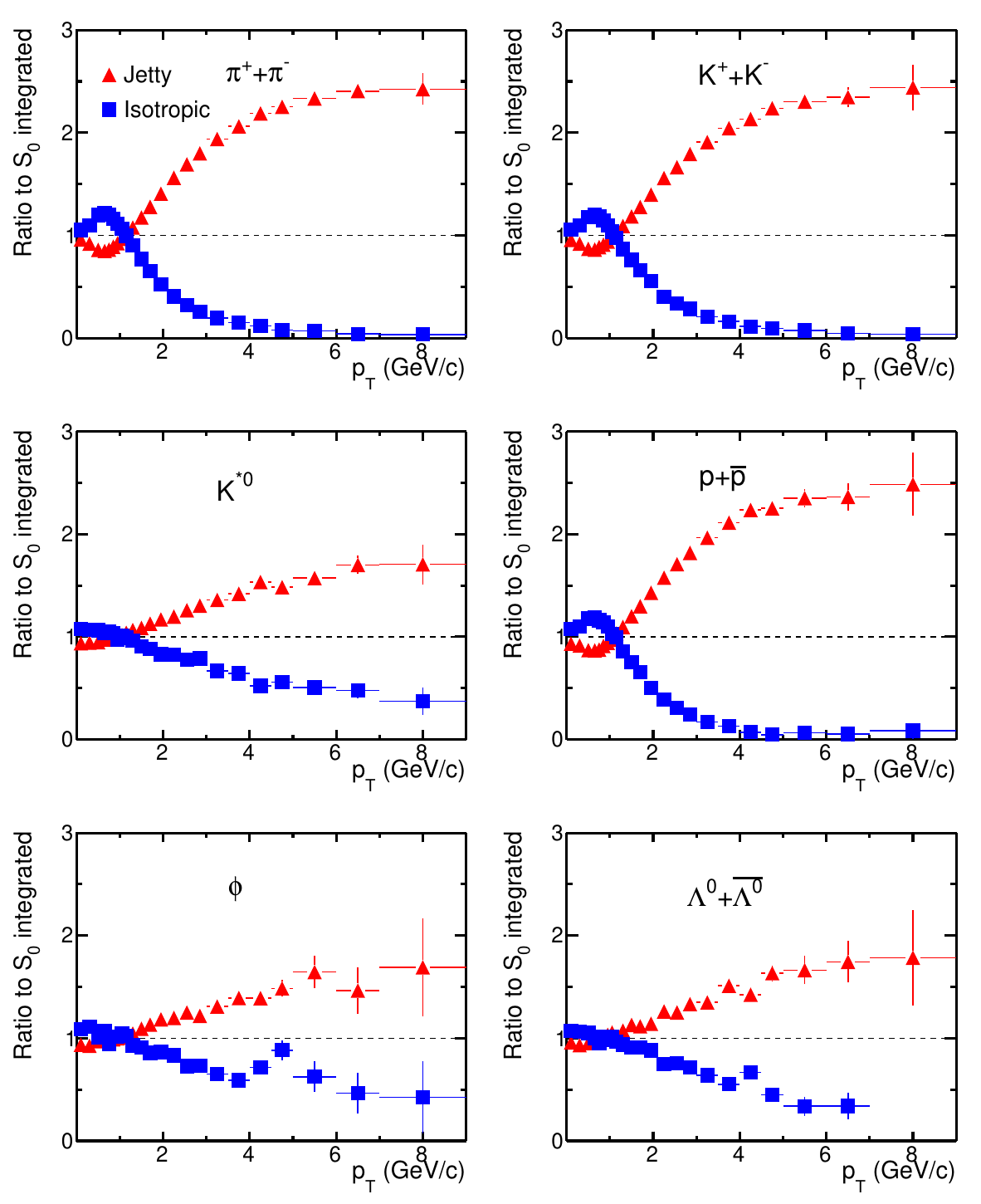}
\caption[]{(Color online) Ratio of $p_{\rm{T}}$-spectra for isotropic and jetty events to the spherocity integrated events for V0M-I (left) and V0M-X (right) multiplicity classes using PYTHIA8. The blue squares are for isotropic events and red triangles are for jetty events.}
\label{ptspectra_sp_V0M_I_X}
\end{figure*}


Figure.~\ref{ptspectra_mult} shows the $p_{\rm{T}}$-spectra for pion, kaon, ${\rm{K}}^{*0}$, proton, $\phi$ and $\Lambda$ at mid-rapidity ($ |\eta|< 0.5$) for different multiplicity classes in pp collisions at $\sqrt{s}=13\rm{~TeV}$. We observe a clear multiplicity dependence of the spectral shapes. As we move from low to high charged particle multiplicity, we observe hardening of $p_{\rm{T}}$-spectra while the bulk production is similar for all the multiplicity classes. This trend is similar to that in experimental data from ALICE~\cite{Acharya:2018orn} for multiplicity dependence study in pp collisions at $\sqrt{s}$ = 7 TeV. 

Left panel of Fig.~\ref{ptspectra_sp_min_bias} shows the spherocity dependence of $p_{\rm{T}}$-spectra of identified particles for minimum bias pp collisions. The right panel shows the ratio of $p_{\rm{T}}$-spectra for isotropic and jetty events to the spherocity integrated events. We observe the crossing of the ratios for pions, kaons and protons at around 3~GeV/c while for other particles we do not observe the crossing of ratios till 6~GeV/c. This suggests that for minimum bias pp collisions, the production of pions, kaons and protons at low $p_{\rm{T}}$ are dominated by isotropic events while after 3~GeV/c, the production is dominated by jetty events. For resonances (${\rm{K}}^{*0}$ and $\phi$) and $\Lambda$, the production is dominated by isotropic events till 6~GeV/c. This indicates different production mechanisms for resonances and heavier particles compared to pions, kaons and protons.

Figure~\ref{ptspectra_sp_V0M_I_X} shows the ratio of $p_{\rm{T}}$-spectra for isotropic and jetty events to the spherocity integrated events for V0M-I (left) and V0M-X (right) multiplicity classes. To see the effect of multiplicity on the crossing point of jetty and isotropic events, we have taken the lowest (V0M-X) and the highest (V0M-I) multiplicity classes for a comparison. 
For stable particles like, $\pi$, K and p, the crossing point moves towards high-$p_{\rm{T}}$, while going from low ($\sim$ 1 GeV/c) to high-multiplicity classes ($\sim$ 3 GeV/c). This indicates that particle production in high-multiplicity collisions are dominated by isotropic events whereas in low-multiplicity it is dominated by jetty events. The preliminary results from ALICE~\cite{Bencedi:2018ctm} show a mass dependence of the crossing points for high-multiplicity pp collisions, which could be attributed to flow-like behaviour. Although the color reconnection in PYTHIA mimics a flow-like behaviour~\cite{Ortiz:2013yxa}, we do not observe mass dependence of crossing points with the default color reconnection setting. 
 Similar to the results from minimum bias pp collisions, we observe crossing points for resonances and $\Lambda$ at higher-$p_{\rm{T}}$ (around 5 GeV/c) for V0M-I class.
 However, we observe the crossing point for low-multiplicity pp collisions (V0M-X class) to be similar for all particles, i.e. around 1 GeV/c.

\subsection{Integrated yields (dN/dy)}
\begin{figure}[ht]
\includegraphics[width=7.5cm, height=7.0cm]{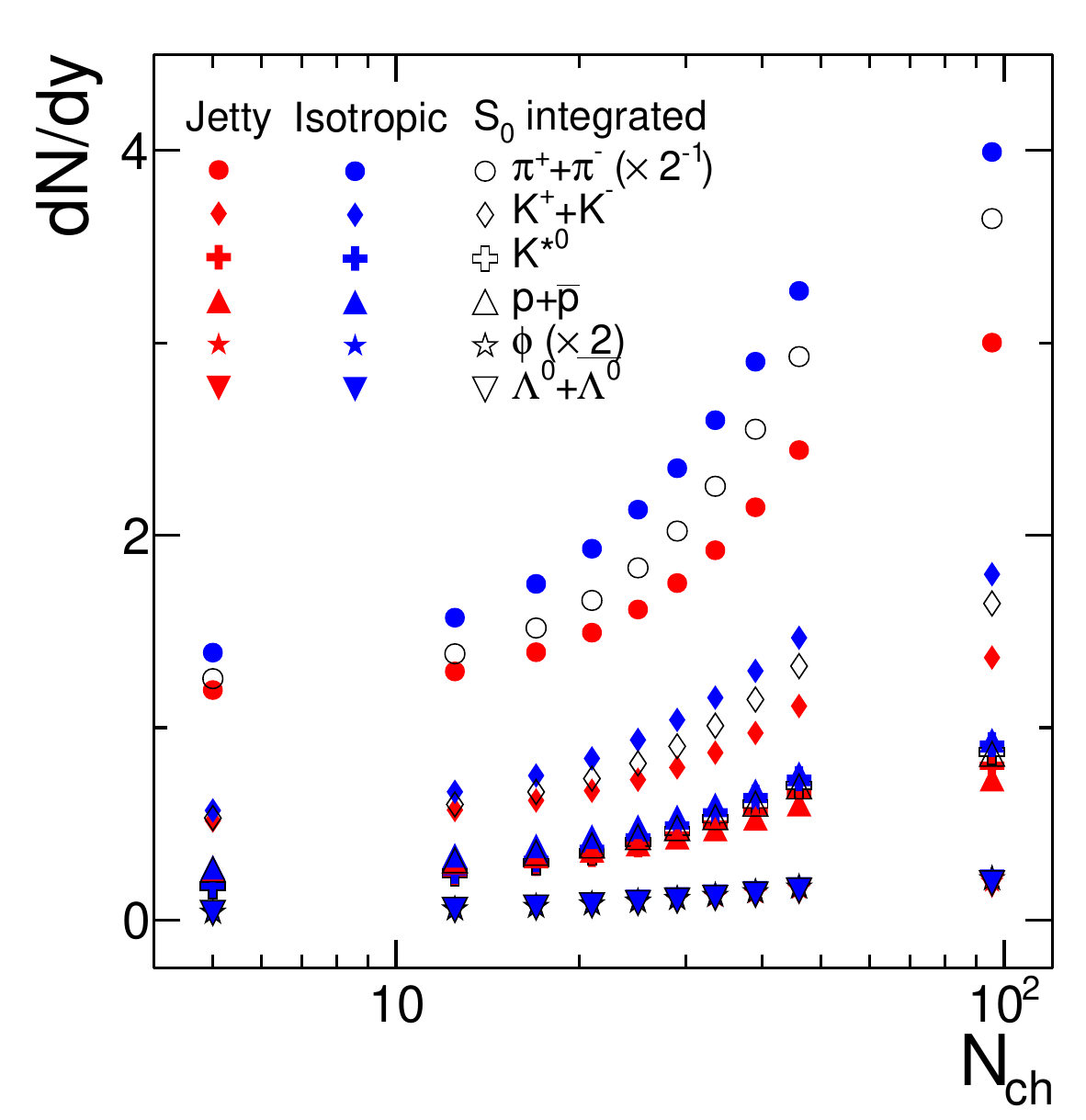}
\includegraphics[width=7.5cm, height=7.0cm]{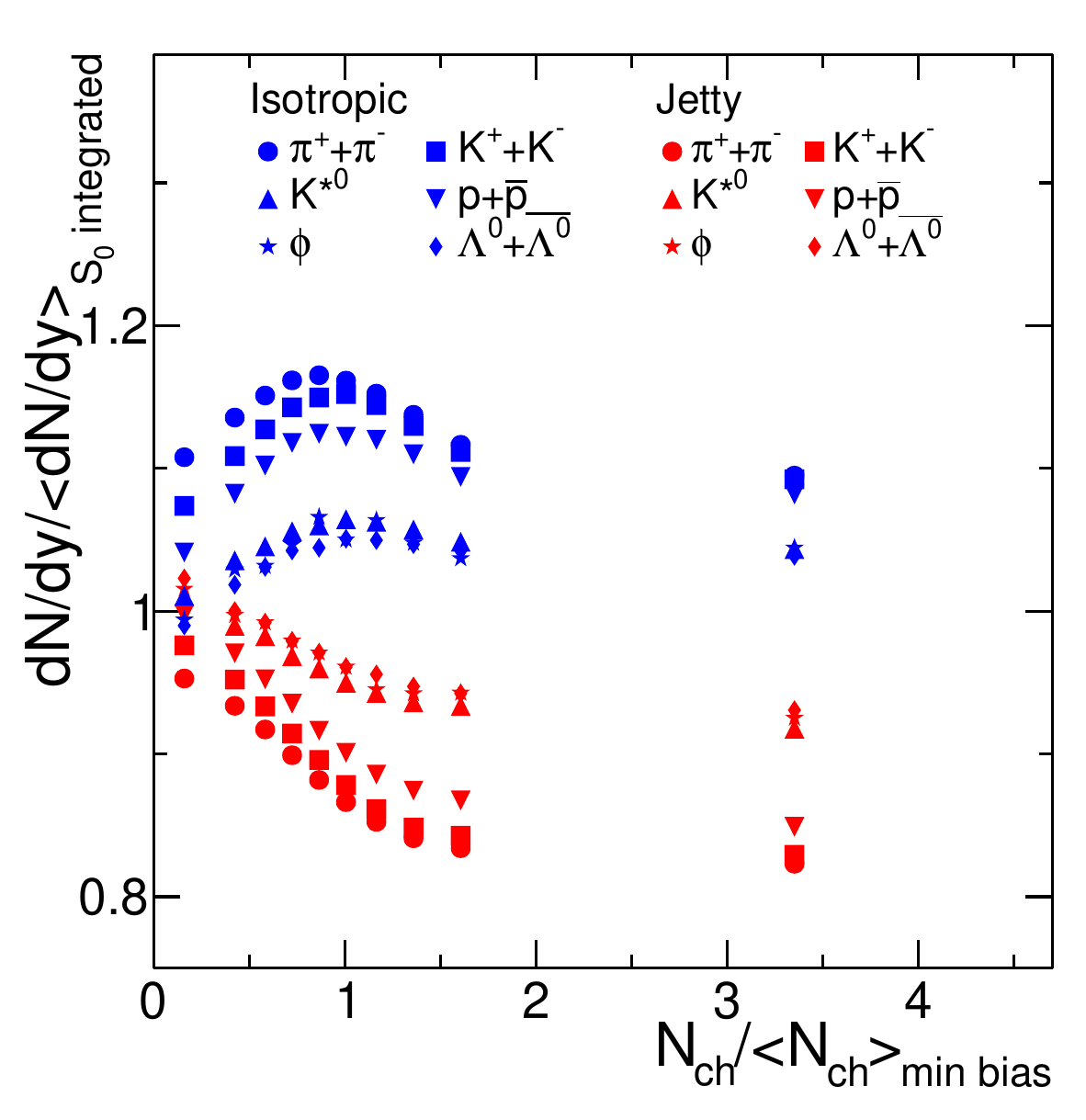}
\caption[]{(Color online) Top: Integrated yield (dN/dy) of pions, kaons, ${K}^{*0}$, protons, $\phi$, and $\Lambda$ at mid-rapidity ($ |\eta|< 0.5$) as a function of charged particle multiplicity for isotropic (blue squares), jetty (red triangles) and spherocity integrated (open circles) events using PYTHIA8. Bottom: The spherocity-integrated-scaled yield showing a separation of jetty and isotropic events towards high-multiplicity classes in pp collisions.}
\label{int_yield}
\end{figure}

Figure~\ref{int_yield} shows the dN/dy of pions, kaons, ${\rm{K}}^{*0}$, protons, $\phi$, and $\Lambda$ at mid-rapidity ($ |\eta|< 0.5$) as a function of charged particle multiplicity for isotropic, jetty and spherocity integrated events for pp collisions at $\sqrt{s}$ = 13 TeV. Pion being the lightest particle, the integrated yield is the maximum compared to other particles. The mass and charged particle multiplicity dependence of integrated yield for spherocity integrated events is similar to the experimental data from ALICE~\cite{Acharya:2018orn} for multiplicity dependence study in pp collisions at $\sqrt{s}$ = 7 TeV.  As it appears, in low-multiplicity classes, the effect of spherocity is minimal, whereas towards higher-multiplicity classes, it starts playing a role in making a separation of jetty to isotropic events for all identified particles. The contribution to integrated yield is always dominated by the isotropic events but we observe significant contributions from jetty events as well. As we have observed in the top panel of Fig.~\ref{sp_dis}, the contribution to dN/dy from jetty events decreases with charged particle multiplicity. The bottom panel of Fig~\ref{int_yield} shows the spherocity integrated-scaled  yield as a function of minimum-bias-scaled charged particle multiplicity. Slowly towards higher multiplicity classes, a clear separation of jetty to isotropic events is observed. However, the separation is the highest for the lightest meson, $\pi$. The values of integrated yield of identified particles for different multiplicity classes in isotropic, jetty and spherocity integrated events are tabulated in Table~\ref{tab:int_yield}.

\subsection{Mean Transverse Momentum ($\langle p_{\rm{T}} \rangle$)}
\begin{figure}[ht]
\includegraphics[width=8.5cm, height=10.5cm]{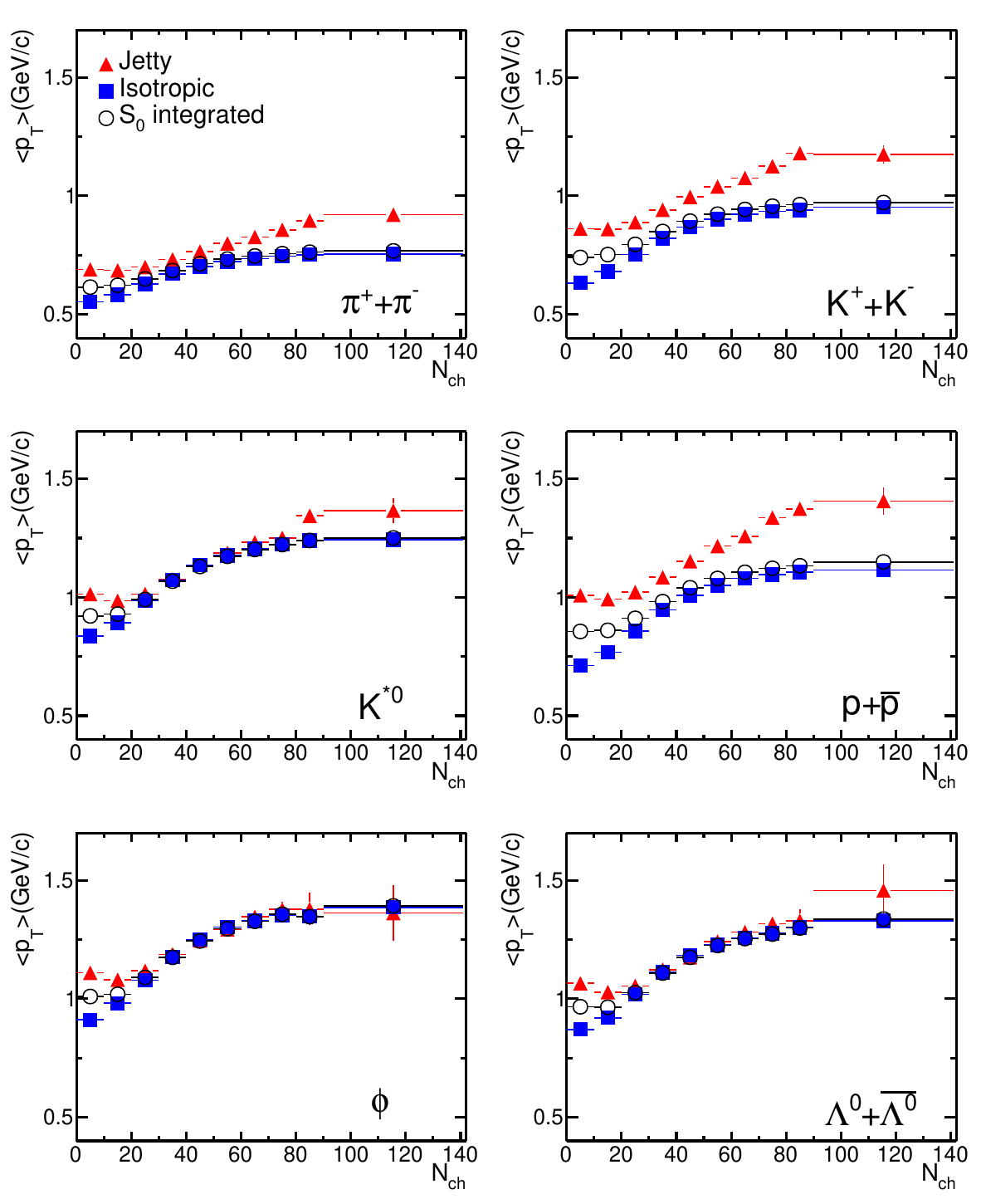}
\caption[]{Mean transverse momentum ($\langle p_{\rm{T}}\rangle$) of pions, kaons, ${K}^{*0}$, protons, $\phi$, and $\Lambda$ at mid-rapidity ($ |\eta|< 0.5$) as a function of charged particle multiplicity for isotropic (blue squares), jetty (red triangles) and spherocity integrated (open circles) events using PYTHIA8.  }
\label{meanpT}
\end{figure}
The $\langle p_{\rm{T}}\rangle$ of pions, kaons, ${\rm{K}}^{*0}$, protons, $\phi$, and $\Lambda$ at mid-rapidity ($ |\eta|< 0.5$) as a function of charged particle multiplicity for isotropic, jetty and spherocity integrated events are shown in Fig.~\ref{meanpT}. The obtained $\langle p_{\rm{T}} \rangle$ for all multiplicity classes in different spherocity events are tabulated in Table~\ref{tab:meanpT} for all the measured particle species. 
For pions, kaons and protons we observe that the $\langle p_{\rm{T}} \rangle$ is higher for jetty events compared to the isotropic events for all the multiplicity classes. This is one of the reasons that one should use spherocity as a selective parameter along with the charged particle multiplicity. However, we see significant differences in $\langle p_{\rm{T}} \rangle$ for other particles at only low-multiplicity classes but the $\langle p_{\rm{T}} \rangle$ remains spherocity independent for high-multiplicity pp collisions for resonances and $\Lambda$.

 After studying these observables in details, let us now focus on the highest multiplicity class, which is of special importance to us for understanding the spherocity dependent $p_{\rm{T}}$-differential particle ratios.

\subsection{Particle Ratios}
\begin{figure}[ht]
\includegraphics[width=8.5cm, height=11.5cm]{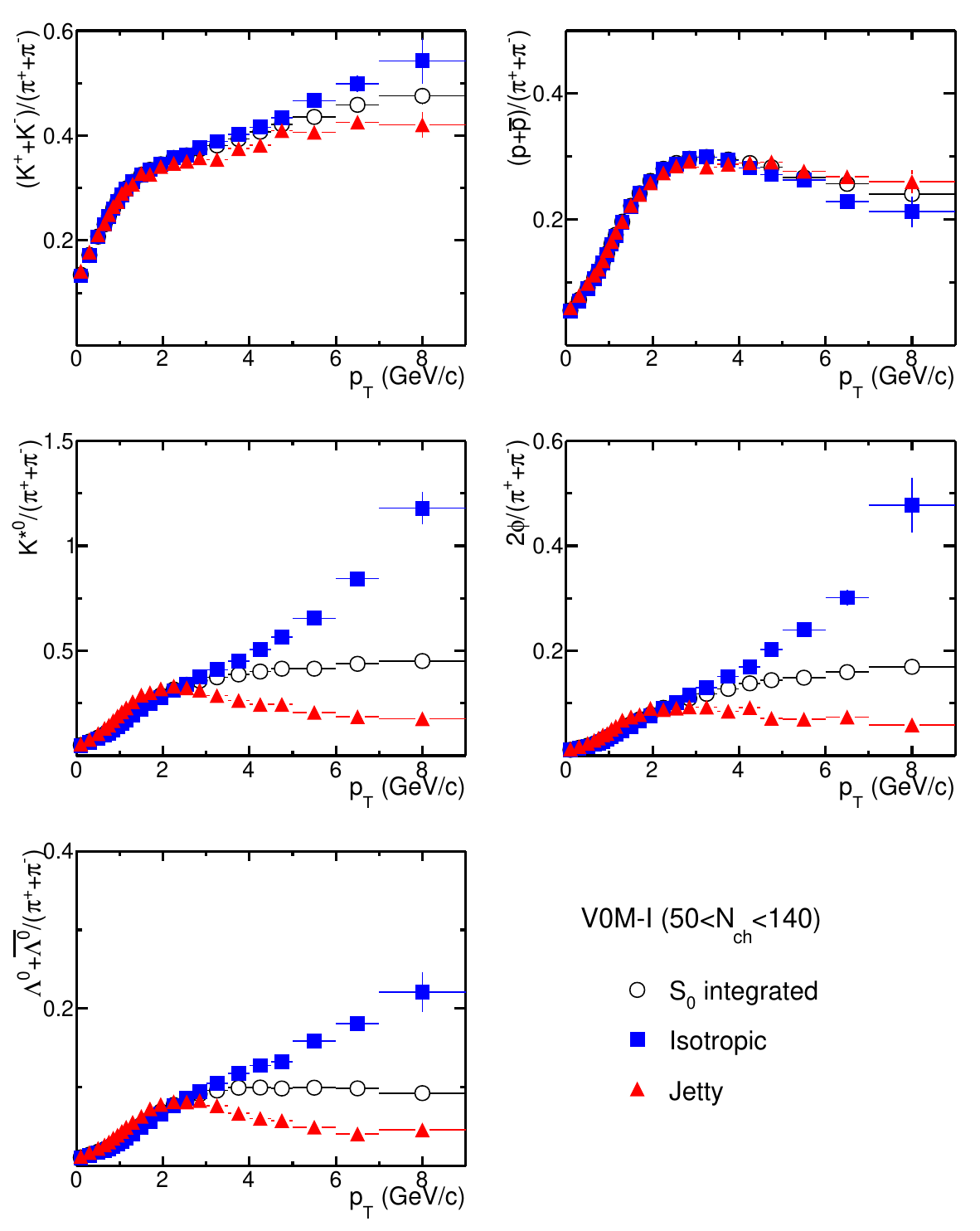}
\caption[]{(Color Online) $p_{\rm{T}}$-differential particle ratio to pions for high-multiplicity pp collisions in isotropic (blue squares), jetty (red triangles) and spherocity integrated (open circles) events using PYTHIA8.}
\label{pT-differential_pion}
\end{figure}

Figure~\ref{pT-differential_pion} shows the $p_{\rm{T}}$-differential particle ratio to pions for high-multiplicity pp collisions in isotropic, jetty and spherocity integrated events. The trend of particle ratios involving K$^{*0}$ and $\phi$ for $S_0$-integrated events are similar to that of experimental data in pp collisions at $\sqrt{s}$ = 13 TeV~\cite{Dash:2018cjh}. The proton to pion ratio seems to be independent of spherocity classes in the discussed $p_{\rm{T}}$-range but we observe a clear spherocity dependence of K, K$^{*0}$, $\phi$ and $\Lambda$ to pion ratio as a function of $p_{\rm{T}}$. At low-$p_{\rm{T}}$, the dependence on spherocity is negligible for all identified particles. However, at high-$p_{\rm{T}}$  the ratios of K, K$^{*0}$, $\phi$ and $\Lambda$ to pion are higher for isotropic events as compared to jetty events.  This gets more pronounced for higher mass particles. As discussed earlier, the average number of MPIs is higher for isotropic events compared to jetty ones. Higher number of MPIs may favour the production of particles with strange quark as the constituents. Our results seem to be consistent with these findings. Further the proton to pion ratio, which is a measure of baryon to meson ratio, is found to be independent of event spherocity. This ratio increases up to $p_{\rm{T}} \sim$ 3-4 GeV/c and then decreases towards higher-$p_{\rm{T}}$ making the high-$p_{\rm{T}}$ domain meson rich. This is similar to the experimental observations \cite{Adam:2015qaa,Abelev:2014laa}. For the ratios of K$^{*0}$, $\phi$ and $\Lambda$ to pions, the spherocity integrated events show a $p_{\rm{T}}$-independent behaviour after around 2 GeV/c, whereas for isotropic events these ratios show a sharp rise. These observations open up new domain of studies, which may shed light on particle production mechanisms taking event multiplicity, spherocity and $p_{\rm{T}}$-differential particle ratios.

\begin{figure}[ht]
\includegraphics[width=8.5cm, height=8.5cm]{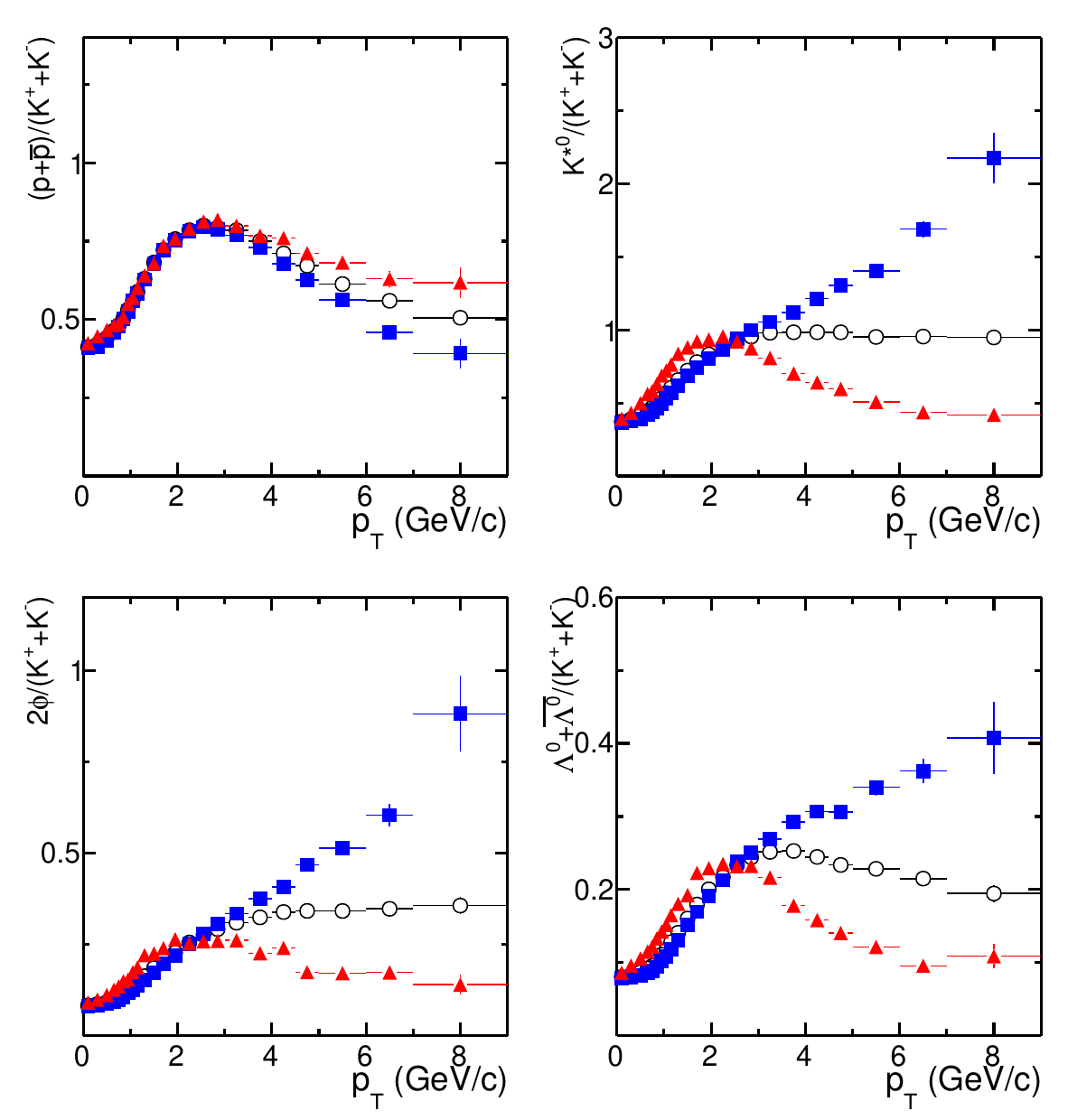}
\caption[]{(Color Online) $p_{\rm{T}}$-differential particle ratio to kaons for high-multiplicity pp collisions in isotropic (blue squares), jetty (red triangles) and spherocity integrated (open circles) events using PYTHIA8.}
\label{pT-differential_kaon}
\end{figure}

The $p_{\rm{T}}$-differential particle ratios to kaons for high-multiplicity pp collisions in isotropic, jetty and spherocity integrated events are shown in Fig.~\ref{pT-differential_kaon}. The proton to kaon ratio seems to be independent of spherocity classes up to $p_{\rm{T}} \sim$ 5 GeV/c. But we observe a clear spherocity dependence of K$^{*0}$, $\phi$ and $\Lambda$ to kaon ratios as a function of $p_{\rm{T}}$. At low-$p_{\rm{T}}$, the K$^{*0}$, $\phi$ and $\Lambda$ to kaon ratios are higher for jetty events and also lower for isotropic events compared to the spherocity-integrated events. However, at high-$p_{\rm{T}}$ a completely opposite behaviour is observed. And this behaviour is similar to particle ratios with respect to pions.

\begin{figure*}[ht]
\includegraphics[width=16cm, height=7cm]{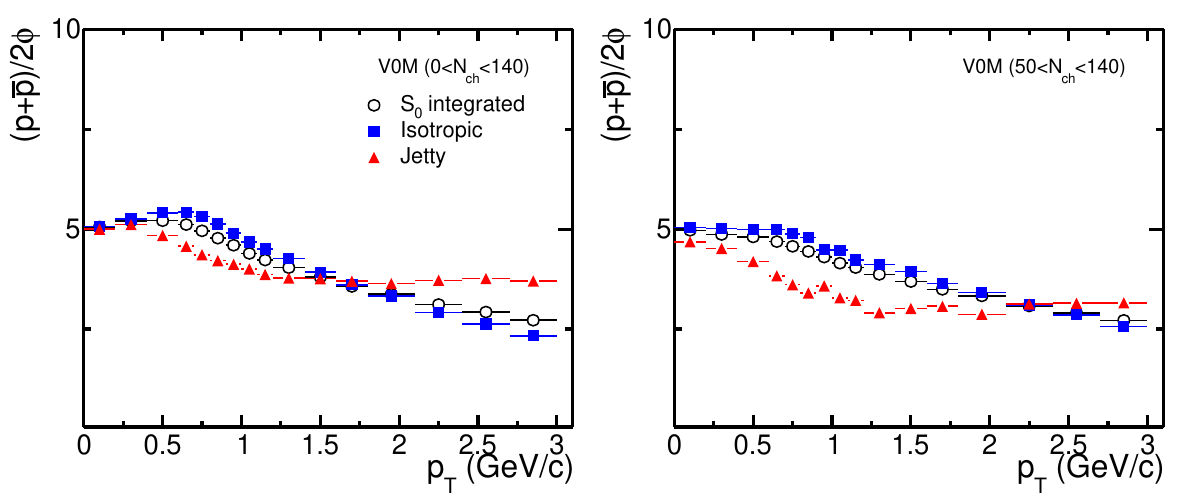}
\caption[]{(Color Online) $p_{\rm{T}}$-differential particle ratio of proton to $\phi$ for minimum bias (left panel) and high-multiplicity (right panel) pp collisions in isotropic (blue squares), jetty (red triangles) and spherocity integrated (open circles) events using PYTHIA8.}
\label{pT-protonTophi}
\end{figure*}

Figure~\ref{pT-protonTophi} shows the $p_{\rm{T}}$-differential particle ratio of proton to $\phi$ for minimum bias (left panel) and high-multiplicity (right panel) pp collisions in isotropic, jetty and spherocity integrated events. Assuming QGP-like behaviour is driven by hydrodynamics, one would expect similar shape of $p_{\rm{T}}$ spectra at low-$p_{\rm{T}}$ for proton and $\phi$ due to their similar masses~\cite{Abelev:2014uua}. A comparison of p/$\phi$ ratio in minimum bias and high-multiplicity pp collisions shows a flatness at low-$p_{\rm{T}}$ in isotropic events for high-multiplicity pp collisions. This hints to QGP-like behaviour in high-multiplicity pp collisions, which might be attributed to CR in PYTHIA8. 

\begin{figure}[h!]
\includegraphics[width=8cm, height=8.5cm]{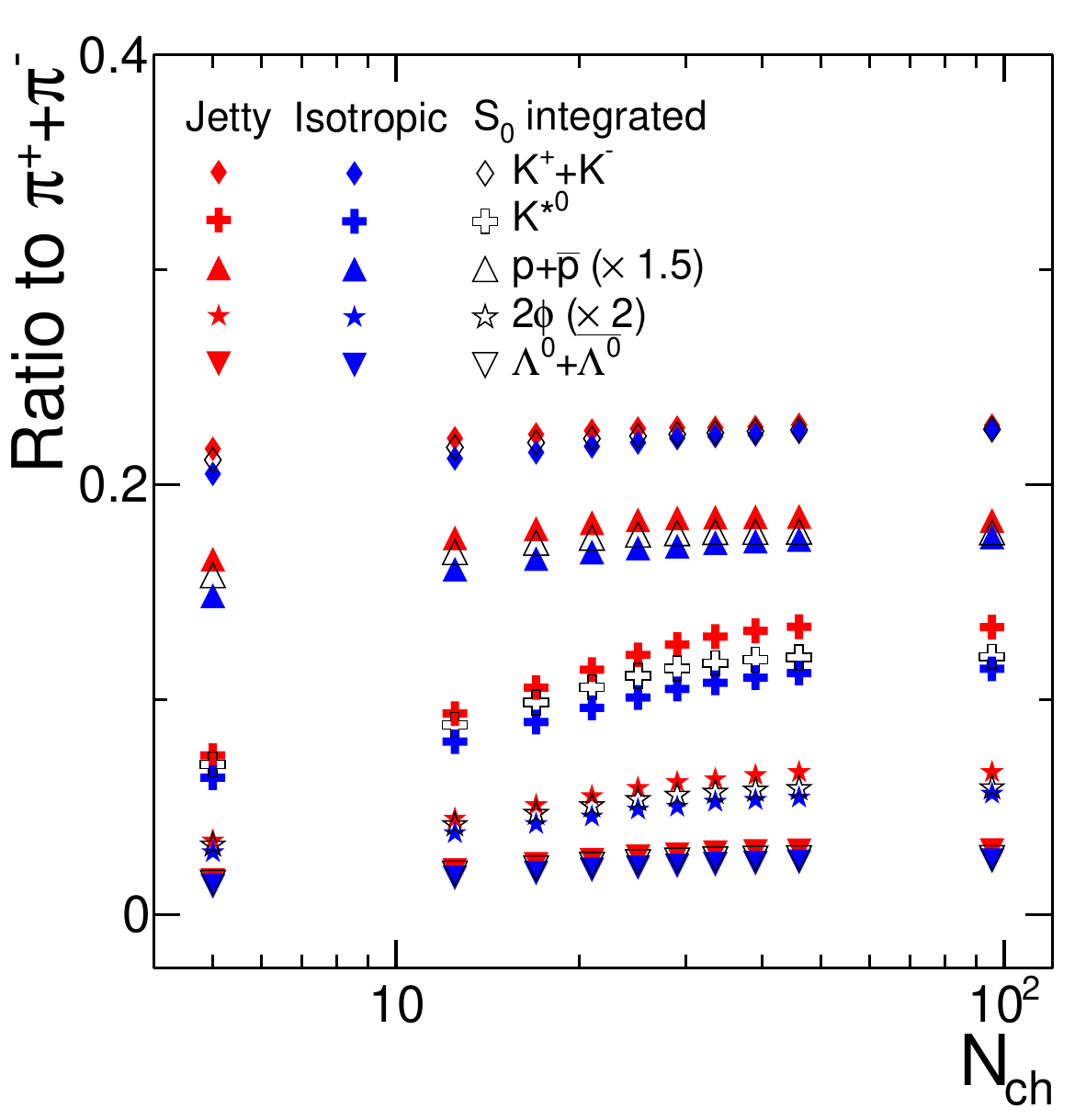}
\includegraphics[width=8cm, height=8.5cm]{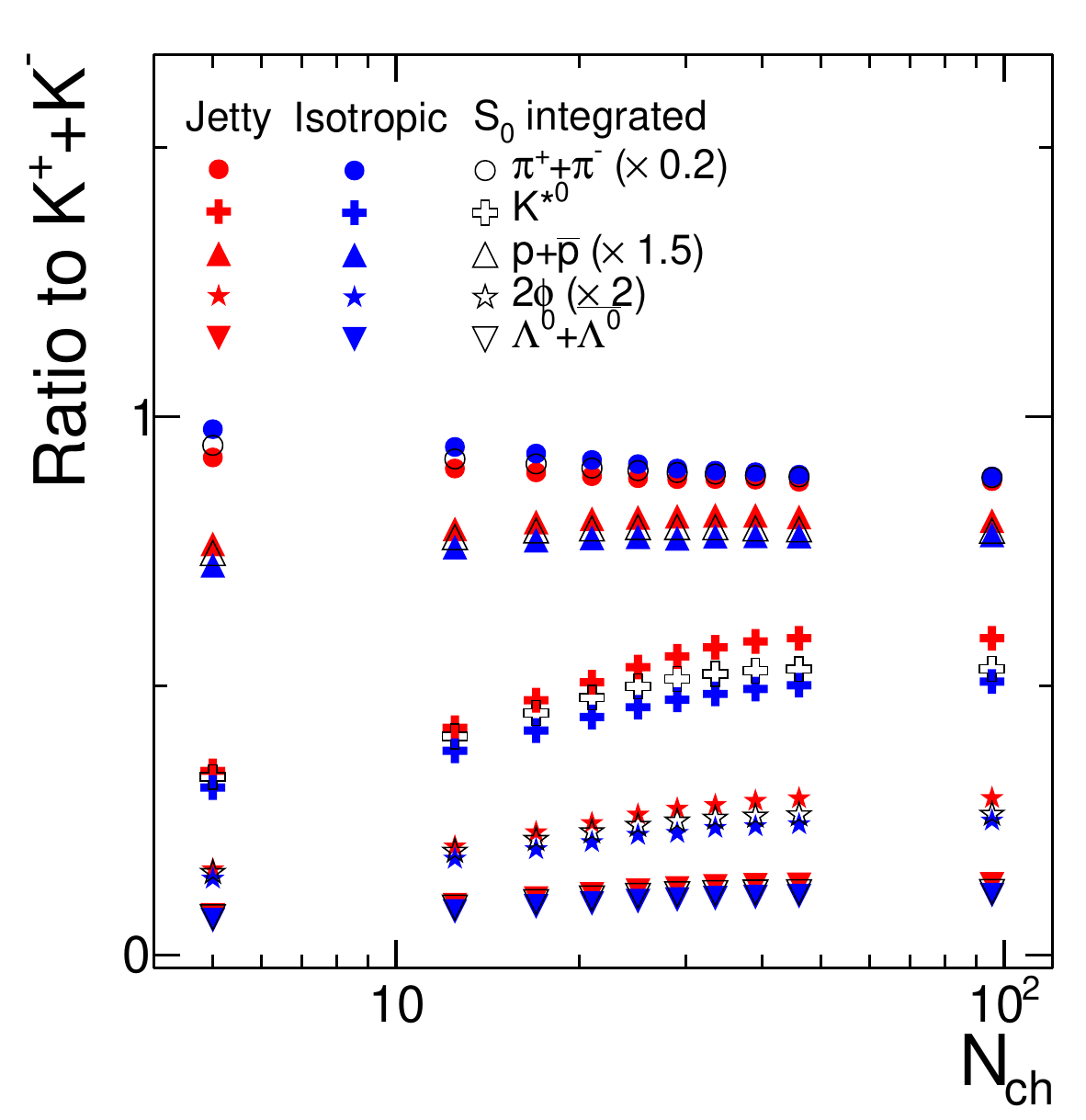}
\caption[]{(Color Online) $p_{\rm{T}}$-integrated particle ratio to pions (upper panel) and kaons (lower panel) for high-multiplicity pp collisions in isotropic (blue squares), jetty (red triangles) and spherocity integrated (open circles) events using PYTHIA8.}
\label{pT-int_pion_kaon}
\end{figure}

 To have a direct comparison to experimental measurements, let us now consider the $p_{\rm{T}}$-integrated particle ratios with respect to pions and kaons as a function of spherocity classes and charged particle multiplicity.
Upper (lower) panel of Fig.~\ref{pT-int_pion_kaon} shows $p_{\rm{T}}$-integrated particle ratios to pions (kaons) for high-multiplicity pp collisions in isotropic, jetty and spherocity integrated events. Although, we observe significant spherocity dependence of integrated yield of different particles (Fig.~\ref{int_yield}), we do not observe  any large dependence of $p_{\rm{T}}$-integrated particle ratios to pions and kaons. This suggests that the relative increase in integrated yield for different particles as a function of spherocity are similar, while indicating that the domain of transverse momentum is very crucial for the nature of particle production. As expected, the ratios of particles with respect to lighter mass particle seem to increase with charged particle multiplicity. A slightly increasing trend of K/$\pi$, K*$^{0}$/$\pi$ and $\Lambda/\pi$ suggests strangeness enhancement in high-multiplicity pp collisions~\cite{ALICE:2017jyt}. The trend of $p_{\rm{T}}$-integrated particle ratios for $\phi$ to $\pi$ and $\phi$ to K in $S_0$-integrated events is similar to that of experimental data in pp collisions at $\sqrt{s}$ = 13 TeV~\cite{Tripathy:2018ehz}. Due to possible re-scattering effects, the experimental data for K*$^{0}$/K decreases with charged particle multiplicity~\cite{Tripathy:2018ehz}. A similar trend is not observed in PYTHIA, as it does not include possible re-scattering effects. While comparing the upper and lower panel of Fig.~\ref{pT-int_pion_kaon}, one can observe that although the spherocity dependence on the particle ratio is not high but the dependence seems to be opposite compared to integrated yield in Fig.~\ref{int_yield}. For all the identified particle ratios to pions and kaons, the contribution from jetty events are higher compared to the isotropic ones. This leads to the conclusion that while studying the QGP-like behaviour in pp collisions, one should separate the jetty events from all the events.

\section{Summary and Conclusion}
\label{summary}
We have made an extensive study of various observables taking spherocity and event multiplicity as key parameters in pp collisions at $\sqrt{s}$ = 13 TeV using the pQCD inspired PYTHIA8 model.  The aim of the present study is to understand the high-multiplicity pp events at the highest LHC energy in view of a possible formation of QGP-droplet in pp collisions. In view of QGP-like behaviours observed in the LHC pp collisions and the fact that pp no longer serves as a baseline to understand a possible nuclear medium formation in heavy-ion collisions at the LHC energies, it is crucial to understand the mechanism of particle production in these high-multiplicity events. Through these simulation studies, we invoke the explicit inclusion of MPI and CR effects responsible for particle production in pp collisions and then through transverse spherocity and charged particle multiplicity dependent analysis we try to understand various aspects of identified particle production in LHC pp collisions. Our findings are summarized as below:

\begin{enumerate}
\item We study the correlation between the multi-partonic interactions, event shape (transverse spherocity) and charged particle multiplicity. The observed results are the interplay of different underlying models in PYTHIA such as SFM, MPI and CR.

\item We report the simulation results on the event shape and charged particle multiplicity dependent study of ($\pi^{+}+\pi^{-}$), (K$^{+}$+K$^{-}$), (p+$\mathrm{\bar{p}}$), K*$^{0}$, $\phi$ and ($\Lambda^{0}+\bar{\Lambda}^{0}$) production in pp collisions at $\sqrt{s}$ = 13 TeV using PYTHIA 8.235 for the first time. This could be confronted to experimental data, when they become available.

\item We explore the event shape dependence of the transverse momentum ($p_{\rm{T}}$) spectra, integrated yield, mean transverse momentum ($\langle p_{\rm{T}} \rangle$) and particle ratios of the identified particles. A clear spherocity dependence of $p_{\rm{T}}$-spectra is observed for all the particles. 

\item The crossing of the ratios of jetty and isotropic events to the spherocity-integrated ones, depend on the multiplicity classes. 

\item $\langle p_{\rm{T}} \rangle$ of ($\pi^{+}+\pi^{-}$), (K$^{+}$+K$^{-}$) and (p+$\mathrm{\bar{p}}$) depend on spherocity while for other particles, $\langle p_{\rm{T}} \rangle$ does not depend on spherocity except for low-multiplicity classes.

\item Contrary to $\langle p_{\rm{T}} \rangle$, the $p_{\rm{T}}$-differential particle ratios to ($\pi^{+}+\pi^{-}$) for K$^{*0}$, $\phi$ and $\Lambda^{0}+\bar{\Lambda}^{0}$ depend on spherocity while the ratios remain almost independent for kaons and protons.

\item Weak $p_{\rm{T}}$-dependence of proton to $\phi$ ratio at low-$p_{\rm{T}}$ in isotropic high-multiplicity pp events indicates a hydrodynamic behaviour. This, along with a signal of strangeness enhancement, hints for QGP-like behaviour in high-multiplicity pp collisions. This behaviour in PYTHIA8 could be due to the CR mechanism.

\item Larger dependence of integrated yield on spherocity is observed for high-multiplicity compared to the low-multiplicity pp collisions. However, the $p_{\rm{T}}$-integrated particle ratio shows less dependence on spherocity, which suggests that the relative increase in integrated yield for different particles as a function of spherocity are similar. 

\item The $p_{\rm{T}}$-differential ratios with respect to pions show enhancement for particles containing strange quark(s). 
This suggests that higher number of MPIs may favour the production of particles with strange quark as the constituent.

\end{enumerate}
The present studies in view of the high-multiplicity era of pp collisions at the LHC energies bear high-level of importance in view of a possible QGP-like medium formation in pp collisions, while giving enough differential information about particle production mechanism taking into account event multiplicity, event spherocity, the transverse momentum and multi-partonic interactions. Availability of experimental data in near future would help us having a better understanding of underlying physics behind the high-multiplicity pp collisions at the LHC energies. In addition, these studies should help in fine-tuning the pQCD based event generators, while comparing similar findings in experimental data.

\section*{Acknowledgements}
The authors acknowledge the financial supports  from  ALICE  Project  No. SR/MF/PS-01/2014-IITI(G)  of  
Department  of  Science  \&  Technology,  Government of India. ST acknowledges the financial support by 
DST-INSPIRE program of Government of India. RS acknowledges the financial supports 
under the DAE-BRNS Project No: 58/14/29/2019-BRNS. The authors gratefully acknowledge the initial discussions with Dr. Antonio Ortiz Velasquez and Dr. Sudipan De. The authors would like to acknowledge the usage of resources  of the LHC grid computing facility at VECC, Kolkata.


\section{Appendix}
\begin{table*}[h]
\caption{Integrated yield of identified particles in isotropic, jetty and $S_{0}$ integrated events.} 
\centering 
\scalebox{0.8}{
\begin{tabular}{cccccccc} 
\hline\hline 
V0M & Event Class & $\pi$ & K & $K^{*0}$ & p & $\phi$ & $\Lambda $ \\
\\ [0.5ex]
\hline 
 & Jetty &6.003		&1.365		&0.802	&0.733	&0.099	&0.178\\[1ex]
{I} &Isotropic &7.985		&1.798		&0.912	&0.935		&0.112	&0.199\\[1ex]
& $S_{0}$ integrated & 7.293		&1.646		&0.874		&0.864	&0.107	&0.192\\[1ex]
\hline
& Jetty &4.886		&1.112	&0.654	&0.603	&0.081	&0.144\\[1ex]
{II} &Isotropic &6.541		&1.467	&0.734	&0.761	&0.089	&0.160\\[1ex]
& $S_{0}$ integrated &5.858		&1.320	&0.701	&0.695	&0.086	&0.153\\[1ex]
\hline
& Jetty &4.292		&0.973	&0.566	&0.529	&0.070	&0.125\\[1ex]
{III} &Isotropic &5.805		&1.295	&0.639	&0.672	&0.077	&0.139\\[1ex]
& $S_{0}$ integrated &5.103		&1.147	&0.605	&0.605	&0.074	&0.132\\[1ex]
\hline
& Jetty &3.845		&0.871	&0.497	&0.474	&0.061	&0.110\\[1ex]
{IV} &Isotropic &5.197		&1.156	&0.560	&0.599	&0.068	&0.121\\[1ex]
& $S_{0}$ integrated &4.510		&1.010	&0.527	&0.535	&0.064	&0.115\\[1ex]
\hline
& Jetty &3.504		&0.793	&0.440	&0.431	&0.054	&0.097\\[1ex]
{V} &Isotropic &4.698		&1.040	&0.493	&0.536	&0.059	&0.106\\[1ex]
& $S_{0}$ integrated &4.044		&0.903	&0.463	&0.478	&0.056	&0.101\\[1ex]
\hline
& Jetty &3.229		&0.730	&0.390	&0.395	&0.048	&0.086\\[1ex]
{VI} &Isotropic &4.267		&0.936	&0.431	&0.485	&0.052	&0.093\\[1ex]
& $S_{0}$ integrated &3.662		&0.814	&0.406		&0.431	&0.049	&0.089\\[1ex]
\hline
& Jetty &2.988		&0.672	&0.340	&0.363	&0.041	&0.075\\[1ex]
{VII} &Isotropic &3.861		&0.840	&0.371	&0.434	&0.044	&0.080\\[1ex]
& $S_{0}$ integrated &3.323		&0.735	&0.351	&0.388	&0.042	&0.077\\[1ex]
\hline
& Jetty &2.785		&0.622	&0.294	&0.333	&0.035	&0.065\\[1ex]
{VIII} &Isotropic &3.495		&0.751	&0.313	&0.386	&0.037	&0.067\\[1ex]
& $S_{0}$ integrated &3.036		&0.666	&0.299	&0.350	&0.036	&0.065\\[1ex]
\hline
& Jetty &2.585		&0.572	&0.241	&0.302	&0.029	&0.053\\[1ex]
{IX} &Isotropic &3.144		&0.667	&0.253	&0.336	&0.030	&0.054\\[1ex]
& $S_{0}$ integrated &2.768		&0.601	&0.244	&0.311	&0.029	&0.053\\[1ex]
\hline
& Jetty &2.392		&0.518	&0.177	&0.263	&0.021	&0.038\\[1ex]
{X} &Isotropic &2.781		&0.570	&0.177	&0.275	&0.020	&0.036\\[1ex]
& $S_{0}$ integrated &2.510		&0.531	&0.175	&0.264	&0.020	&0.037\\[1ex]
\hline 
\end{tabular}
}
\label{tab:int_yield}
\end{table*}

\begin{table*}[h]
\caption{Average transverse momentum of identified particles in GeV/c for isotopic, jetty and $S_{0}$ integrated events.} 
\centering 
\scalebox{0.8}{
\begin{tabular}{cccccccc} 
\hline\hline 
V0M & Event Class & $\pi$ & K & $K^{*0}$ & p & $\phi$ & $\Lambda $ \\
\\ [0.5ex]
\hline 
 & Jetty &0.811	&1.056	&1.205		&1.237	&1.312	&1.258\\[1ex]
{I} &Isotropic &0.731	&0.913	&1.193	&1.067	&1.319	&1.245\\[1ex]
& $S_{0}$ integrated &0.740	&0.934		&1.189		&1.093	&1.313	&1.241\\[1ex]
\hline
& Jetty &0.769	&1.001	&1.142	&1.161	&1.251	&1.183\\[1ex]
{II} &Isotropic &0.705	&0.873	&1.142	&1.014	&1.258	&1.190\\[1ex]
& $S_{0}$ integrated & 0.716	&0.897	&1.137	&1.046	&1.251	&1.182\\[1ex]
\hline
& Jetty &0.747	&0.966	&1.105	&1.114		&1.218	&1.148\\[1ex]
{III} &Isotropic &0.685	&0.843	&1.101	&0.976	&1.204	&1.145\\[1ex]
& $S_{0}$ integrated & 0.698	&0.870	&1.097	&1.009	&1.205	&1.139\\[1ex]
\hline
& Jetty &0.728		&0.935	&1.070	&1.077	&1.182	&1.117\\[1ex]
{IV} &Isotropic &0.665	&0.812	&1.062		&0.934	&1.167	&1.098\\[1ex]
& $S_{0}$ integrated &0.680	&0.843	&1.058	&0.972	&1.164	&1.097\\[1ex]
\hline
& Jetty &0.712	&0.909	&1.039	&1.046	&1.136	&1.074\\[1ex]
{V} &Isotropic &0.646	&0.780	&1.020	&0.893	&1.118	&1.051\\[1ex]
& $S_{0}$ integrated &0.664	&0.818	&1.022		&0.939	&1.124		&1.056\\[1ex]
\hline
& Jetty &0.700	&0.890	&1.013		&1.024	&1.124	&1.056\\[1ex]
{VI} &Isotropic &0.626	&0.750	&0.983	&0.855	&1.078	&1.021\\[1ex]
& $S_{0}$ integrated &0.649		&0.795	&0.989	&0.911	&1.090	&1.026\\[1ex]
\hline
& Jetty &0.690	&0.870	&0.995	&1.000	&1.092	&1.035\\[1ex]
{VII} &Isotropic &0.607	&0.719	&0.942	&0.815	&1.034	&0.970\\[1ex]
& $S_{0}$ integrated &0.635	&0.772	&0.958	&0.884	&1.052	&0.991\\[1ex]
\hline
& Jetty &0.683	&0.858	&0.982	&0.986	&1.071	&1.022\\[1ex]
{VIII} &Isotropic &0.589	&0.691	&0.904	&0.778	&0.989	&0.930\\[1ex]
& $S_{0}$ integrated &0.624	&0.755	&0.932	&0.862	&1.021	&0.966\\[1ex]
\hline
& Jetty &0.687	&0.860	&0.988	&0.996	&1.088	&1.034\\[1ex]
{IX} &Isotropic &0.572	&0.662	&0.866	&0.744	&0.949	&0.895\\[1ex]
& $S_{0}$ integrated &0.619		&0.746	&0.920	&0.855	&1.009		&0.957\\[1ex]
\hline
& Jetty &0.689	&0.861	&1.013	&1.008		&1.109	&1.065\\[1ex]
{X} &Isotropic &0.553	&0.633	&0.836	&0.712	&0.911	&0.871\\[1ex]
& $S_{0}$ integrated &0.615	&0.740	&0.922	&0.856	&1.009	&0.966\\[1ex]
\hline 
\end{tabular}
}
\label{tab:meanpT}
\end{table*}


\begin{thebibliography}{99}

\bibitem{ALICE:2017jyt} 
  J.~Adam {\it et al.} [ALICE Collaboration],
  Nature Phys.\  {\bf 13}, 535 (2017).
  
  \bibitem{Mangano:2017plv} 
  M.~L.~Mangano and B.~Nachman,
  Eur.\ Phys.\ J.\ C {\bf 78}, 343 (2018).
  
\bibitem{Bzdak:2014dia} 
  A.~Bzdak and G.~L.~Ma,
  Phys.\ Rev.\ Lett.\  {\bf 113}, 252301 (2014).
  
  \bibitem{Bozek:2013ska} 
  P.~Bozek, W.~Broniowski and G.~Torrieri,
  Phys.\ Rev.\ Lett.\  {\bf 111}, 172303 (2013).
    
  \bibitem{Dumitru:2010iy} 
  A.~Dumitru, K.~Dusling, F.~Gelis, J.~Jalilian-Marian, T.~Lappi and R.~Venugopalan,
  Phys.\ Lett.\ B {\bf 697}, 21 (2011).
  
  \bibitem{Ma:2014pva} 
  G.~L.~Ma and A.~Bzdak,
  Phys.\ Lett.\ B {\bf 739}, 209 (2014).
  
  \bibitem{Sjostrand:2007gs} 
  T.~Sjostrand, S.~Mrenna and P.~Z.~Skands,
  Comput.\ Phys.\ Commun.\  {\bf 178}, 852 (2008).
   
  \bibitem{Sjostrand:2013cya} 
  T.~Sjöstrand,
  arXiv:1310.8073 [hep-ph].
  
    \bibitem{Ortiz:2013yxa} 		
  A.~Ortiz Velasquez, P.~Christiansen, E.~Cuautle Flores, I.~Maldonado Cervantes and G.~Paic,
  Phys.\ Rev.\ Lett.\  {\bf 111}, 042001 (2013).
  
  \bibitem{Bierlich:2015rha} 
  C.~Bierlich and J.~R.~Christiansen,
  Phys.\ Rev.\ D {\bf 92}, 094010 (2015).
  
  \bibitem{Acconcia:2017bjv} 
  R.~Acconcia, D.~D.~Chinellato, R.~Derradi de Souza, J.~Takahashi, G.~Torrieri and C.~Markert,
  Phys.\ Rev.\ D {\bf 97}, 036010 (2018).
   
  \bibitem{Sjostrand:2006za} 
  T.~Sjostrand, S.~Mrenna and P.~Z.~Skands,
  JHEP {\bf 0605}, 026 (2006).
  
  \bibitem{Abazov:2009gc} 
  V.~M.~Abazov {\it et al.} [D0 Collaboration],
  Phys.\ Rev.\ D {\bf 81}, 052012 (2010).
  
  \bibitem{Chekanov:2007ab} 
  S.~Chekanov {\it et al.} [ZEUS Collaboration],
  Nucl.\ Phys.\ B {\bf 792}, 1 (2008).
  
  \bibitem{Abelev:2012rz} 
  B.~Abelev {\it et al.} [ALICE Collaboration],
  Phys.\ Lett.\ B {\bf 712}, 165 (2012).
  
  \bibitem{Thakur:2017kpv} 
  D.~Thakur, S.~De, R.~Sahoo and S.~Dansana,
  Phys.\ Rev.\ D {\bf 97}, 094002 (2018).
  
  \bibitem{Abelev:2012hy} 
  B.~Abelev {\it et al.} [ALICE Collaboration],
  Eur.\ Phys.\ J.\ C {\bf 72}, 2183 (2012).
  
  \bibitem{Aamodt:2011zza} 
  K.~Aamodt {\it et al.} [ALICE Collaboration],
  Eur.\ Phys.\ J.\ C {\bf 71}, 1594 (2011).
  
  \bibitem{Abelev:2012sk} 
  B.~Abelev {\it et al.} [ALICE Collaboration],
  Eur.\ Phys.\ J.\ C {\bf 72}, 2124 (2012).
  
   \bibitem{Cuautle:2014yda} 
E.~Cuautle, R.~Jimenez, I.~Maldonado, A.~Ortiz, G.~Paic and E.~Perez, 
  arXiv:1404.2372 [hep-ph].
  
    \bibitem{Cuautle:2015kra} 
  A.~Ortiz, G.~Paic and E.~Cuautle,
  Nucl.\ Phys.\ A {\bf 941}, 78 (2015).

  \bibitem{Bencedi:2018ctm} 
  G.~Bencedi [ALICE Collaboration], Nucl. Phys. A \textbf{982}, 507 (2019).
  
  \bibitem{pythia8html}
  Pythia8 online manual:(http://home.thep.lu.se/~torbjorn/pythia81html/Welcome.html).

  \bibitem{Skands:2014pea} 
  P.~Skands, S.~Carrazza and J.~Rojo, Eur. Phys. J. C \textbf{74}, 3024 (2014) and CERN-PH-TH-2014-069.

  \bibitem{Abelev:2014ffa} 
  B.~B.~Abelev {\it et al.} [ALICE Collaboration],
  Int.\ J.\ Mod.\ Phys.\ A {\bf 29}, 1430044 (2014). 
  
  \bibitem{Ortiz:2017jho} A.~Ortiz, 
Adv. Ser. Direct. High Energy Phys. \textbf{29}, 343 (2018).
  
  \bibitem{Salam:2009jx} G.~P.~Salam, 
  Eur.\ Phys.\ J.\ C {\bf 67}, 637 (2010).    
     
      \bibitem{Adam:2015pza} 
  J.~Adam \textit{et al.} [ALICE Collaboration], 
  Phys. Lett. B \textbf{753}, 319 (2016).
     
  \bibitem{Acharya:2018orn} 
  S.~Acharya {\it et al.} [ALICE Collaboration],
  Phys. Rev. C \textbf{99}, 024906 (2019).
  
  \bibitem{Dash:2018cjh} 
  A.~K.~Dash [ALICE Collaboration],
  Nucl. Phys. A \textbf{982}, 467 (2019).
  
  \bibitem{Adam:2015qaa} 
  J.~Adam {\it et al.} [ALICE Collaboration],
  Eur.\ Phys.\ J.\ C {\bf 75}, 226 (2015).
  
  \bibitem{Abelev:2014laa} 
  B.~B.~Abelev {\it et al.} [ALICE Collaboration],
  Phys.\ Lett.\ B {\bf 736}, 196 (2014).
  
  \bibitem{Abelev:2014uua} 
  B.~B.~Abelev {\it et al.} [ALICE Collaboration],
  Phys.\ Rev.\ C {\bf 91}, 024609 (2015).
  
  \bibitem{Adam:2017zbf} J.~Adam {\it et al.} [ALICE Collaboration], 
  Phys.\ Rev.\ C {\bf 95}, 064606 (2017).
  
  \bibitem{Tripathy:2018ehz} 
  S.~Tripathy [ALICE Collaboration],
  Nucl. Phys. A \textbf{982}, 180 (2019).
    
  
\end{thebibliography}
\end{document}